\documentclass{article} 

\usepackage{amssymb} 
\usepackage{latexsym,amssymb} 
\usepackage{graphics} 
\usepackage{pstricks,psfrag} 
\usepackage{amsmath} 
\usepackage{subfigure} 
\usepackage{vmargin} 
\usepackage{epsfig} 

\begin{document} 

\large 

\begin{titlepage} 

\title{\bf Flavour physics of the RS model with KK masses reachable at LHC} 

\vskip 1cm 

\author{G. Moreau \footnote{e-mail: {\tt greg@cftp.ist.utl.pt}} , 
J. I. Silva-Marcos \footnote{e-mail: {\tt juca@cftp.ist.utl.pt}} \\ \\ 
{\it  CFTP, Departamento de F\'{\i}sica} \\ 
{\it  Instituto Superior T\'ecnico, 
Avenida Rovisco Pais, 1} \\ 
{\it 1049-001 Lisboa, Portugal}} 
\maketitle 

\vskip 1cm 

\begin{abstract} 
The version of the higher-dimensional Randall-Sundrum (RS) model with matter in the bulk, 
which addresses the gauge hierarchy problem, has additional attractive features. 
In particular, it provides an intrinsic geometrical mechanism 
that can explain the origin of the large mass hierarchies among the Standard Model fermions. 
Within this context, a good solution for the gauge hierarchy problem 
corresponds to low masses for the Kaluza-Klein (KK) excitations of the gauge bosons. 
Some scenarios have been proposed in order to render these low masses 
(down to a few TeV) consistent with precision electroweak measurements. 
Here, we give specific and complete realizations of this RS version with small KK masses, 
down to 1 TeV, which are consistent with the entire structure of the fermions in flavour 
space: (1) all the last experimental data on quark/lepton masses and 
mixing angles (including massive neutrinos of Dirac type) are reproduced, (2) flavour 
changing neutral current constraints are satisfied and (3) the effective suppression 
scales of non-renormalizable interactions (in the physical basis) are within the bounds 
set by low energy flavour phenomenology. Our result, on the possibility of having KK gauge 
boson modes as light as a few TeV, constitutes one of the first theoretical motivations 
for experimental searches of direct signatures at the LHC collider, of this interesting 
version of the RS model which accommodates fermion masses. 
\end{abstract} 

\vskip 1cm 

PACS numbers: 11.10.Kk, 12.15.Ff, 14.60.St 

\end{titlepage} 

\section{Introduction} 
\label{intro} 

The old idea of the existence of additional spatial dimensions \cite{Kaluza,Klein},   
a fundamental ingredient for string theories, has recently received a 
renewed interest due to several proposals for universal extra dimension 
models \cite{UED} (with all Standard Model fields propagating in extra 
dimensions), brane universe models \cite{Akama}-\cite{RS} (with 
Standard Model fields living in our 3-dimensional spatial subspace) and 
intermediate models \cite{IntermA,IntermB,IntermC} (with gauge and Higgs 
bosons in the bulk, fermions being confined at fixed points along 
extra dimensions). 

Amongst the brane models, the one suggested by Randall and Sundrum (RS) 
\cite{Gog,RS}, and its different extensions, has attracted particular 
attention. A considerable advantage of the RS scenario is that it addresses 
the so-called gauge hierarchy problem (i.e. the huge discrepancy between the 
gravitational and the electroweak scale) without introducing any new energy scale 
in the fundamental theory. 

In a variant of the original RS set-up, which matured over the years 
\cite{GNeubertA}-\cite{hep-th/9912232}, all the Standard Model (SM) particles 
except the Higgs boson (to ensure that the gauge hierarchy problem does not re-emerge) 
have been promoted to bulk fields rather than brane fields. 
\\ This RS version possesses the three following important phenomenological characteristics. 
First, unification of the gauge couplings at high scale is possible within a Grand 
Unified Theory (GUT) framework \cite{UNI-RS}-\cite{UNI-bulk}. 
Secondly, from the cosmological point of view, there exists a viable Kaluza-Klein 
(KK) WIMP candidate \cite{KKrelic} for the dark matter of universe \cite{LZP2,LZP1}. 
Finally, this RS version provides a totally new physical interpretation 
\cite{RSloc,RSmany} for the origin of the large mass hierarchy prevailing 
among all different flavours and types of SM fermions 
\footnote{Within the RS context, other higher-dimensional mechanisms 
\cite{GNeubertA,MultiBrNu,Appel,GherAl} apply specifically to neutrinos in 
order to explain their lightness compared to the rest of SM fermions.}. 
Such an interpretation of the whole SM fermion mass hierarchy is attractive, 
as it does not rely on the presence of any new symmetry in the short-distance 
theory, in contrast with the usual Froggatt-Nielsen mechanism \cite{FN} 
where a flavour symmetry is required. As a matter of fact, this interpretation 
is purely geometrical: it is based on the possibility of different localizations 
for SM fermions along extra dimension, depending on their flavour and 
type \footnote{This possibility of fermion localizations along extra dimension(s) 
was also considered in the context of large flat extra dimension models, in order 
to generate quark \cite{Branco} and lepton \cite{Frere} masses/mixings.}. 
In such a scenario, the quark masses and CKM mixing angles can be effectively 
accommodated \cite{HSquark,Hflav,Chang}, as well as the lepton masses and MNS 
mixing angles in both cases where neutrinos acquire Majorana masses (via either 
dimension five operators \cite{HSHHLL} or the see-saw mechanism \cite{HSseesaw}) 
and Dirac masses (see \cite{HSpre}, and, \cite{MnuPred} for order unity Yukawa couplings 
leading to mass hierarchies essentially generated by the higher-dimensional mechanism). 

In the present article, we will elaborate concrete, complete and coherent realizations 
of the RS scenario, with bulk matter, which simultaneously: 
{\it (i)} reproduce all the last experimental data on quark/lepton masses and mixings   
(in the minimal case of Dirac neutrino masses where right handed neutrinos are added 
to SM fields) through the above geometrical mechanism   
{\it (ii)} satisfy the strongest Flavour Changing Neutral Current (FCNC) constraints 
(FCNC effects will be calculated, including new ones) for masses of the first KK gauge 
boson excitation down to $m_{KK}=1 \mbox{TeV}$ 
\footnote{In our notation, $m_{KK}=m_{KK}^{(1)}(\gamma)=m_{KK}^{(1)}(gluon)$ 
is the mass of first KK excitation of the photon and gluon.} 
{\it (iii)} generate non-renormalizable operator scales in agreement with low energy 
phenomenology (a realistic analysis in the physical basis will be performed)   
{\it (iv)} respect all the remaining constraints, e.g. the intrinsic 
theoretical bound on the curvature of the $AdS_5$ space. 

In preliminary works \cite{Hflav,HSpre}, where realizations of the RS scenario (with SM bulk fields) 
were constructed in order to simultaneously create correct SM fermion masses/mixings and 
obey the FCNC constraints, the mass of first KK excitation of gauge bosons was taken to be high: 
$m_{KK}=10 \mbox{TeV}$. In this way, the FCNC effects, due to the 
significant flavour dependence of fermion locations (needed to generate SM fermion mass 
hierarchies), were suppressed because in fact they are conveyed by the exchange of KK states 
of the gauge bosons (see Section \ref{FCNC} for details). 
Here, in contrast, we will show that the data on SM fermion masses/mixings can 
be compatible with FCNC bounds for $m_{KK}$ values down to $1 \mbox{TeV}$ 
\footnote{The possibility of some FCNC signatures of the RS model, with bulk matter and 
KK masses around 3 $\mbox{TeV}$, was discussed for the B-physics \cite{Soni} as well as in 
rare K and top decays \cite{Perez}.}. 
\\ Our result, of a conceivable light KK gauge boson excitation ($1 \mbox{TeV}$), 
is important in the sense that it reopens the possibility, for the attractive 
version of the RS model generating SM fermion masses, to be tested at the Large Hadron Collider (LHC) 
\cite{LHC} with a centre-of-mass energy at $14 \mbox{TeV}$. 
For such a test to be possible, a scenario should apply relaxing the severe electroweak (EW) 
precision constraints, e.g. the ones proposed in \cite{LR} or \cite{EWBa,EWBb,EWF}. 
Assuming the scenario in \cite{LR} with a left-right gauge structure, 
one can expect to obtain some signatures of the RS model at LHC 
via KK gauge boson exchanges, since these KK states can be as light as 
$\sim 3 \mbox{TeV}$ (limit from EW bounds). 
\\ Moreover, this result, i.e. the possibility of having low KK gauge field masses, is in favour of a good 
solution for the gauge hierarchy problem. Indeed, lower KK masses correspond typically to an effective 
gravity scale on our brane which is closer to the electroweak scale. 

The organization of this paper is as follows. In Section \ref{Hmass}, we describe consistent realizations 
of the RS scenario which generate the correct SM fermion mass hierarchies. Then in Section \ref{FCNC}, 
the FCNC effects appearing in these RS realizations are computed and we show that those fulfil well 
the relevant experimental constraints for $m_{KK}$ values down to $1 \mbox{TeV}$. Our method to obtain 
so small $m_{KK}$ values remaining acceptable is also exposed there. Other experimental constraints, 
like those coming from precision EW data, are discussed in Section \ref{other}. In Section \ref{NR}, 
we calculate the effective suppression scales of four dimensional operators in physical basis and 
demonstrate that, within the above RS realizations, the suppression scale values induce FCNC process 
amplitudes in agreement with experimental bounds. Finally, in Section \ref{conclu}, we summarize and 
discuss the impacts of our results. 

\section{Generation of mass hierarchies} 
\label{Hmass} 

\subsection{The RS set-up} 
\label{setup} 

We consider the RS scenario with all SM fields residing in the bulk, except the 
Higgs boson which is confined on the TeV-brane (see below). 
Recall that the RS scenario consists of a 5-dimensional theory where the extra 
spatial dimension (denoted by $y$) is compactified over a $S^{1}/\mathbb{Z}_{2}$ 
orbifold with radius $R_{c}$ ($-\pi R_{c}\leq y \leq \pi R_{c}$). Gravity also 
propagates inside the bulk and the extra dimension is bordered by two 3-branes 
with tensions $\Lambda_{(y=0,\pi R_{c})}$ (vacuum energy densities) tuned such 
that, 
\begin{equation} 
\Lambda_{(y=0)}=-\Lambda _{(y=\pi R_{c})}=-\Lambda /k=24kM_{5}^{3}, 
\label{RStensions} 
\end{equation} 
$\Lambda$ being the bulk cosmological constant, $M_{5}$ the fundamental 
5-dimensional gravity scale and $k$ a RS characteristic energy scale (see 
below). Within this framework, there exists a solution to the 5-dimensional 
Einstein's equations respecting 4-dimensional Poincar\'{e} invariance. 
It corresponds to a zero mode of the graviton localized on the positive 
tension brane (3-brane at $y=0$) and to the non-factorisable metric: 
\begin{equation} 
ds^{2}=e^{-2\sigma (y)}\eta _{\mu \nu }dx^{\mu }dx^{\nu }+dy^{2},\ \mbox{with}\ 
\sigma (y)=k|y|,   
\label{RSmetric} 
\end{equation} 
where $x^{\mu}\ [\mu =1,\dots ,4]$ are the coordinates for the familiar 4 
dimensions and $\eta _{\mu \nu }=diag(-1,1,1,1)$ is the 4-dimensional flat 
metric. The bulk geometry, associated to this metric, is a 
slice of Anti-de-Sitter ($AdS_5$) space with curvature radius $1/k$. 

We now discuss the energy scales that will be considered. While 
on the brane at $y=0$ (Planck-brane) the effective gravity 
scale is equal to the (reduced) Planck mass: $M_{Pl}=1/\sqrt{8\pi G_{N}}=2.44\ 
10^{18}\mbox{GeV}$ ($G_{N}\equiv$ Newton constant), on the other brane at 
$y=\pi R_{c}$ (TeV-brane) the gravity scale, 
\begin{equation} 
M_{\star }=w\ M_{Pl},   
\label{RSratio} 
\end{equation} 
is suppressed by the exponential `warp' factor $w=e^{-\pi kR_{c}}$. 
We see from Eq.(\ref{RSratio}) that for a small extra dimension 
$R_{c}\simeq 11/k$ (the $k$ value being typically close to $M_5$), 
one finds $w\sim 10^{-15}$ so that $M_{\star}={\cal O}(1)\mbox{TeV}$, 
thus solving the gauge hierarchy problem. 
For these values of $R_c$, $M_5$ is close to the effective 
4-dimensional gravity scale $M_{Pl}$: 
\begin{equation} 
M_{Pl}^{2}=\frac{M_{5}^{3}}{k}(1-e^{-2\pi kR_{c}}).   
\label{RSkrelat} 
\end{equation} 
>From the phenomenological point of view, each one of the models in \cite{LR,EWBa,EWBb,EWF}, 
designed for softening the constraints from EW precision data, permits a value for $m_{KK}$ 
that can be as small as $\sim 3 \mbox{TeV}$. Hence, one can take a maximal $m_{KK}$ value 
of $10 \mbox{TeV}$. This value corresponds to: 
\begin{equation} 
kR_{c}=10.11 
\label{kRvalue} 
\end{equation} 
Indeed, the maximal value of $m_{KK}$ ($m_{KK}=2.45 \ k \ e^{-k \pi R_c}$) is determined by 
the $kR_{c}$ value and the theoretical bound (guarantying that the solution for the metric can 
be trusted) on the 5-dimensional curvature scalar $R_5$, 
\begin{equation} 
|R_5|=|-20 k^2|<M_5^2, 
\label{Rbound} 
\end{equation} 
which, together with relation (\ref{RSkrelat}), leads to: $k<0.105 \ M_{Pl}$. 
\\ The value chosen in Eq.(\ref{kRvalue}) gives rise to the effective gravity 
scale: $M_{\star}=39.2 \mbox{TeV}$ (see Eq.(\ref{RSratio})). This energy scale 
is close to the electroweak symmetry breaking scale even if it is not exactly 
identical. Furthermore, in the context of model in reference \cite{LR} with a left-right 
gauge structure, the needed fine tuning on Higgs boson mass (having a dominant 
loop contribution coming from KK mode exchanges) is estimated to be of the order 
of $1\%$ in the mass-squared. 

\subsection{SM fermion masses and mixings} 
\label{values} 

\noindent $\bullet$ {\bf 5-dimensional masses:} 
In order to produce the SM fermion mass hierarchies via the higher-dimensional 
mechanism mentioned in Section \ref{intro}, the SM (zero mode) fermions must 
possess different localizations along the extra dimension. 
Hence, each type of SM fermion field $\Psi _{i}$ ($i=\{1,2,3\}$ being the family 
index) is coupled to a distinct 5-dimensional mass $m_{i}$ in the fundamental theory: 
\begin{equation} 
\int d^{4}x\int dy\ \sqrt{G}\ m_{i}\bar{\Psi}_{i}\Psi _{i},  \label{5Dmass} 
\end{equation} 
where $G=e^{-8\sigma (y)}$ ($\sigma (y)$ is defined in Eq.(\ref{RSmetric})) 
is the determinant of the RS metric (see \cite{Camb} for another mechanism of fermion 
confinement along the extra dimension). A necessary condition to modify the location 
of SM fermions is that the masses $m_{i}$ have a non-trivial dependence on 
the fifth dimension, more precisely a `kink' profile \cite{Rubakov,Rebbi}. 
For example, these masses could be vacuum expectation values of some scalar 
fields. An attractive possibility is to take \cite{Tamvakis}: 
\begin{equation} 
m_{i}=c_{i}\ \frac{d\sigma (y)}{dy}=\pm \ c_{i}\ k,   
\label{VEV} 
\end{equation} 
the $c_{i}$ being new dimensionless parameters (note that this structure for the mass   
does not conflict with the $\mathbb{Z}_{2}$ symmetry of the $S^{1}/\mathbb{Z}_{2}$ 
orbifold). Then the 5-dimensional fermion fields decompose as ($n$ labelling the 
tower of KK excitations), 
\begin{equation} 
\Psi _{i}(x^{\mu },y)=\frac{1}{\sqrt{2\pi R_{c}}}\sum_{n=0}^{\infty }\psi 
_{i}^{(n)}(x^{\mu })f_{n}^{i}(y),  \label{dec} 
\end{equation} 
admitting the following solution for the zero mode wave function along extra 
dimension \cite{GNeubertA,RSloc}, 
\begin{equation} 
f_{0}^{i}(y)=\frac{e^{(2-c_{i})\sigma (y)}}{N_{0}^{i}},  \label{0mode} 
\end{equation} 
where the normalization factor reads as, 
\begin{equation} 
N_{0}^{i\ 2}=\frac{e^{2\pi kR_{c}(1/2-c_{i})}-1}{2\pi kR_{c}(1/2-c_{i})}. 
\label{Norm} 
\end{equation} 
>From Eq.(\ref{0mode}), we see that when $c_{i}$ increases the zero mode 
of associated fermion gets more localized toward the Planck-brane. 

\noindent $\bullet$ {\bf Mass matrices:} 
In the present framework, as in the SM, fermions acquire a Dirac mass 
through a Yukawa coupling to the Higgs boson. This coupling reads 
as (starting from the 5-dimensional action), 
\begin{equation} 
\int d^4x \int dy \ \sqrt{G} \ \bigg ( Y_{ij}^{(5)} \ H \bar \Psi_{+ i} 
\Psi_{- j} + h.c. \bigg ) = \int d^4x \ M_{ij} \ \bar \psi_{L i}^{(0)} 
\psi_{R j}^{(0)} + h.c. + \dots  \label{Yuk} 
\end{equation} 
The $Y_{ij}^{(5)}$ are the 5-dimensional Yukawa coupling constants 
and the dots stand for KK mass terms. The 
effective 4-dimensional mass matrix is obtained after integrating: 
\begin{equation} 
M_{ij} = \int dy \ \sqrt{G} \ \frac{Y_{ij}^{(5)}}{2 \pi R_c} \ H 
f_0^i(y) f_0^j(y).  \label{MassMatrix} 
\end{equation} 
As discussed in Section \ref{intro} and \ref{setup}, the Higgs profile 
has, typically, a shape peaked at the TeV-brane. Let us assume the 
exponential form: 
\begin{equation} 
H=H_0 \ e^{4k(|y|-\pi R_c)},  \label{Hprofile} 
\end{equation} 
which is motivated by the equation of motion for a bulk scalar field 
\cite{GWise}. Based on the $W^\pm$ boson mass expression, the amplitude $H_0$ 
can be expressed as a function of $kR_c$ and the 5-dimensional weak gauge 
coupling constant $g^{(5)}$. The Yukawa coupling constants are chosen 
almost universal: $Y_{ij}^{(5)}=\kappa_{ij} g^{(5)}$ with $0.9 \leq |\kappa_{ij}| \leq 1.1$, 
following the philosophy adopted for example in \cite{HSquark,HSHHLL,MnuPred}, 
so that the quark/lepton mass hierarchies are mainly governed by the 
geometrical mechanism considered. We assume that the fermion mass matrix in Eq.(\ref{MassMatrix}) 
is real. In order to reproduce CP violating observables, one would have to introduce 
complex phases in Yukawa couplings. For a treatment of CP physics in the RS 
scenario with bulk matter, see \cite{Khalil}-\cite{Perez}. 

Here, we consider the minimal massive neutrino scenario where 
three right handed neutrinos are added to the SM field content so that 
neutrinos acquire Dirac masses. There are no Majorana mass terms for the 
right handed neutrinos as we impose lepton number symmetry. Our motivation for 
imposing lepton number symmetry is to stabilize the proton: as a matter of fact, 
it seems that there exist no quark/lepton localizations which simultaneously 
fit fermion masses and generate effective non-renormalizable operator 
scales inducing an acceptable proton life time \cite{HSquark,Hflav}. 
\\ The explicit expression of effective 4-dimensional Dirac mass matrix 
(\ref{MassMatrix}) was given in \cite{MnuPred}. This mass matrix is only a 
function of $\kappa_{ij}$, $kR_{c}$ and $c_{i}$ parameters. Hence, the dependences 
of down-quark, up-quark, charged lepton and neutrino Dirac mass matrices read 
respectively as, 
\begin{displaymath} 
M_{ij}^{d}=M_{ij}^{d}(\kappa _{ij}^{d},kR_{c},c_{i}^{Q},c_{j}^{d})\ \ 
\mbox{and}\ \ M_{ij}^{u}=M_{ij}^{u}(\kappa _{ij}^{u},kR_{c},c_{i}^{Q},c_{j}^{u}),   
\end{displaymath} 
\begin{equation} 
M_{ij}^{l}=M_{ij}^{l}(\kappa _{ij}^{l},kR_{c},c_{i}^{L},c_{j}^{l})\ \ 
\mbox{and}\ \ M_{ij}^{\nu}=M_{ij}^{\nu}(\kappa _{ij}^{\nu},kR_{c},c_{i}^{L},c_{j}^{\nu}).   
\label{depend} 
\end{equation} 
$\kappa_{ij}^{d,u,l,\nu}$ are associated respectively to the down-quark, up-quark, 
charged lepton and neutrino Yukawa couplings, $c_{j}^{d,u,l,\nu}$ parameterize the 
5-dimensional masses for right handed fermions and $c_{i}^{Q,L}$ for fields belonging 
to quark/lepton $SU(2)_{L}$ doublets. 
\\ For the considered fermion locations (depending on $c_{i}$), 
the mixings between zero modes of quarks/leptons and their first KK modes (localized 
at the TeV-brane), induced by the Yukawa couplings, are insignificant (see \cite{Hflav} 
for details). Indeed, the KK fermion states decouple, for $m_{KK}$ values of the order 
of the TeV scale as chosen here, since their masses (also depending on $c_{i}$ 
\cite{HRizzoBIS}) are larger than $m_{KK}$. As a consequence, the SM fermion 
masses and mixing angles can be reliably calculated from the matrix (\ref{MassMatrix}) 
for the zero modes as the mass corrections due to KK modes are neglectable (even 
at the one loop-level \cite{HSquark,Muon}). Even for the top quark, which 
has the largest wave function overlap with the Higgs boson and thus also with the KK 
quark excitations, these mass corrections are not significant compared to the 
uncertainty on its own mass (see below). Another consequence is that the 
variation of the effective number of neutrinos contributing to the $Z^{0}$ boson width, 
induced by the mixings between the zero and KK states of neutrinos \cite{MultiBrNu}, is 
in agreement with its experimental limit (see \cite{HSpre} for precisions). 

\noindent $\bullet$ {\bf Experimental data:} 
In order to be rigorous, one should specify that the fermion masses 
(\ref{MassMatrix}) are running masses at the cutoff energy scale of 
effective 4-dimensional theory (which is in the TeV range). This is a common 
scale, close to the electroweak scale, at which there is no influence from 
flavour dependent evolution of Yukawa couplings on the fermion mass hierarchy. 
The theoretical predictions for charged lepton masses, quark masses 
and mixing angles, derived from the matrices in Eq.(\ref{depend}), will be 
fitted with the associated values taken at the $Z^0$ boson mass scale 
({\it c.f.} Appendix \ref{appendA}). In order to take into account the 
effect of renormalization group from the $Z^0$ mass scale up to the TeV 
cutoff scale, we assume an uncertainty of $5\%$ 
\footnote{When the error (see Appendix \ref{appendA}) 
on renormalized masses and mixing angles at the 
$Z^0$ mass scale exceeds $5\%$, we admit an uncertainty equal to this error.} 
on the charged lepton masses, quark masses and mixing angles (this 
effect is of a few percent for charged leptons between pole masses and TeV 
scale \cite{hep-ph/9912265}). This significant uncertainty agrees with 
our philosophy of not fixing the fundamental parameters at a high-level 
accuracy. For similar reasons, we will consider the experimental data on neutrino 
masses and lepton mixing angles only at the $4\sigma$ level ({\it c.f.} Appendix 
\ref{appendA}). 
\\ One must also impose the experimental limits on 
absolute neutrino masses. In our case of Dirac neutrino masses, the relevant 
limits are the ones extracted from tritium beta decay experiments ({\it c.f.} 
Appendix \ref{appendA}) which hold irrespective of the nature of neutrino mass. 

\noindent $\bullet$ {\bf Obtained solutions:} 
In Appendix \ref{appendB}, we present 3 points [A,B,C] ([X,Y,Z]) of parameter space 
constituted by $\{\kappa_{ij}^{d,u};c_{k}^{Q,d,u}\}$ ($\{\kappa_{ij}^{l,\nu};c_{k}^{L,l,\nu}\}$) 
which reproduce (via matrices (\ref{depend})) all current experimental data on quark 
(lepton) masses and mixing angles, summarized in Appendix \ref{appendA}, with the accuracy 
discussed in the previous paragraph. 
\\ In fact, for the points A, B and C, the parameter $c_{3}^u$ can lie respectively in the range 
$[0.30,0.35]$, $[0.00,0.15]$ and $[-0.40,-0.08]$ where the quark masses and mixings are 
still reproduced with the allowed accuracy. These ranges correspond to variations of the 
top quark mass inside its uncertainty interval (({\it c.f.} Appendix \ref{appendA}). 
\\ Next we comment on the obtained $c_i$ values in Appendix \ref{appendB}. First, we observe that 
the absolute values of $c_i$ are close to each other. In other terms, for fundamental parameters 
of almost the same order of magnitude, the higher dimensional mechanism generates 
strong hierarchies among the physical quark/lepton masses. This important result means 
that the SM fermion mass hierarchy problem is really addressed. Secondly, all the $|c_i|$ 
values are close to unity, which is desirable for two reasons. The first reason is that 
in this case the 5-dimensional masses $|m_i|$ in Eq.(\ref{VEV}) are close to the scale 
$k$ (being of a similar order as the gravity scale $M_5$). In other words, no other   
energy scale, with a significantly different value, is introduced. Thus the RS model 
maintains its quasi uniqueness of order of magnitude for the various energy scales. 
The other reason is that the values of $|c_i|$ (defined by Eq.(\ref{VEV})) can be chosen 
safely if, 
\begin{equation} 
|c_i| < \sqrt{20}. 
\end{equation} 
This follows from condition $|m_i|<M_5$ and constraint (\ref{Rbound}). 
\\ Concerning the Yukawa coupling constants obtained in Appendix \ref{appendB}, 
we note that a certain accuracy is required for some of them. Nevertheless, 
this accuracy can be lowered by choosing other Yukawa couplings with a higher 
precision. 

\section{FCNC constraints} 
\label{FCNC} 

\subsection{FCNC origin} 
\label{origin} 

Within the SM, there are no FCNC's at tree level, and most loop-induced FCNC 
effects are extremely small. In contrast, 
within the context of RS model with bulk matter, FCNC processes can be induced at 
tree level by exchanges of KK excitations of neutral gauge bosons. Indeed, 
these KK states possess FC couplings as we will discuss now. 

Let us consider 
the action of the effective 4-dimensional coupling between SM fermions $\psi_{i}^{(0)}(x^{\mu})$ 
and KK excitations of any neutral gauge boson $A_{\mu}^{(n)}(x^{\mu})$. In the interaction 
(or weak) basis, it reads as, 
\begin{equation} 
S_{gauge}= g_L^{SM} \int d^4x \sum_{n=1}^{\infty} 
\bar \psi_{Li}^{(0)} \ \gamma^{\mu} \ {\cal C}_{Lij}^{(n)} \ \psi_{Lj}^{(0)} \ A_{\mu}^{(n)} 
\ + \ \{ L \leftrightarrow R \}, 
\label{Sweak} 
\end{equation} 
where $g_{L/R}^{SM}$ is the relevant SM gauge coupling constant and 
${\cal C}_{Lij}^{(n)}$ the $3 \times 3$ diagonal matrix   
$diag(C^{(n)}(c_1),C^{(n)}(c_2),C^{(n)}(c_3))$. The coefficient $C^{(n)}(c_i)$ quantifies 
the wave function overlap along extra dimension between SM fermions ($f_{0}^{i}(y)$) 
and the $n^{th}$ KK mode of the neutral gauge boson. This coefficient corresponds to 
the coefficient $C_{00n}^{f \bar f A}$ in reference \cite{HRizzoBIS}. 
The action in Eq.(\ref{Sweak}) can be rewritten in the mass (or physical) basis (indicated 
by the prime): 
\begin{equation} 
S_{gauge}= g_L^{SM} \int d^4x \sum_{n=1}^{\infty} 
\bar \psi_{L\alpha}^{(0) \ \prime} \ \gamma^{\mu} \ V_{L\alpha\beta}^{(n)} 
\ \psi_{L\beta}^{(0) \ \prime} \ A_{\mu}^{(n)} 
\ + \ \{ L \leftrightarrow R \}, 
\label{Sphysical} 
\end{equation} 
where, 
\begin{equation} 
V_{L}^{(n)}=U_L^\dagger \ {\cal C}_{L}^{(n)} \ U_L, 
\label{Vmatrix} 
\end{equation} 
$U_L$ being the unitary matrix of basis transformation for left handed fermions 
and $\alpha,\beta$ being flavour indexes. In conclusion, the non-universality of the 
effective coupling constants $g_{L/R}^{SM} \times C^{(n)}(c_i)$ between KK modes of 
the gauge fields and the three SM fermion families (which have different 
locations along $y$), in the weak basis, induces non-vanishing off-diagonal 
elements for matrix $V_{L/R}^{(n)}$, in the physical basis, giving rise 
to FC couplings (see Eq.(\ref{Sphysical})). 

\subsection{Small FCNC effects with low KK masses} 
\label{small} 

The mass hierarchies and mixings of SM fermions require different values for the $c_i$ parameters 
(Section \ref{values}), or equivalently different fermion locations, which induce FCNC effects 
at tree level (Section \ref{origin}). These FCNC effects can be suppressed by choosing 
the $c_i$ values within a certain type of configuration, as we will discuss now. Thus, the FCNC bounds 
can be respected even for some low KK masses (FCNC reactions being due to KK mode exchanges). 
\\ The idea is to search for $c_i$ configurations reproducing fermion masses, where the 
$c_i$ parameters, being of a same type (for instance of type $c^d$ or $c^L$) and with different 
values for each generation, are larger than about $0.5$ (for example: $c_1^d=0.7$, 
$c_2^d=0.8$ and $c_3^d=0.9$). Indeed, in this $c_i$ value domain, the coupling constants 
$g_{L/R}^{SM} \times C^{(n)}(c_i)$ are quasi universal among three families since the overlap 
between any KK gauge state and a fermion is almost independent of the fermion localization 
({\it c.f.} \cite{HRizzoBIS} with conventions such that their parameter $\nu$ is identified as 
our $-c$). Therefore, the FC couplings of the KK states of the neutral gauge bosons, 
appearing in the physical basis (see previous subsection), almost vanish. 
\\ With respect to the third family, associated to the heaviest fermion, it is difficult to find a 
configuration where the $c_3$ value for each type of fermion is either similar to the $c_1$ 
and $c_2$ values, or, higher than 
around $0.5$ (heavy fermions should typically correspond to small $c$ values to be localized 
near the TeV-brane where the Higgs field is confined) at the same time as first two $c_1$ and $c_2$   
\footnote{Illustrative examples of this feature are all the values of $c^u_3$ for 
points A,B,C and $c^L_3$ for points Y,Z in Appendix \ref{appendB}.}. 
However, this is compensated by the fact that, for the third fermion generation, FCNC 
constraints are less severe \cite{RSloc} 
\footnote{This comment on the third fermion family can be reformulated from a predictive 
point of view as follows. In RS scenarios generating SM fermion mass hierarchies, fermion locations 
for the third family generally do not correspond to the configurations described in text above. 
Therefore, FCNC effects involving third family are typically larger. This can be observed in various 
tables of this section.}. 

As a matter of fact, all $c_i$ values presented in Appendix \ref{appendB} have been 
obtained accordingly to two main criteria: (I) they reproduce the quark/lepton masses and 
mixing angles, as discussed in previous section (II) they resemble the $c_i$ configurations described 
above. Thus, the six points of parameter space given in Appendix \ref{appendB} verify the 
various experimental FCNC constraints with KK neutral gauge boson masses as low as $m_{KK}=1 \mbox{TeV}$, 
as we are going to show in the following (including FCNC effects induced at one loop level). 

\vspace{0.5cm} 

$\bullet$ {\bf $l_\alpha \to l_\beta l_\gamma l_\gamma$ decay:} 
First, we study the pure leptonic reactions which are free from hadronic uncertainties. 
In the present framework, the FCNC leptonic decay channels for charged leptons $\mu^-$ and $\tau^-$ 
are induced via processes of type $l_\alpha \to l_\beta Z/\gamma^{(n)\star} \to l_\beta l_\gamma l_\delta$ 
(where $_{\alpha,\beta,\gamma,\delta}$ are flavour indexes), i.e. 
by exchanges of virtual KK excitations of the $Z^0$ boson or photon which have FC couplings. 
For instance, the analytical expressions for the widths of these decay channels have been 
calculated in \cite{London}, within a model-independent analysis of constraints 
on new physics (based on effective lagrangian techniques), as a function of the elements of the 
leptonic FC matrix, here denoted by $V_{L/R\alpha\beta}^{(n)}$, in the KK 
gauge field action (\ref{Sphysical}). This matrix $V_{L/R}^{(n)}$ is completely determined for each point 
X,Y,Z of parameter space given in Appendix \ref{appendB}. Indeed, each parameter set X,Y,Z 
fixes the charged lepton mass matrix $M^{l}$ (see Eq.(\ref{depend})) and thus the matrix $U^{l}_{L/R}$ 
(which diagonalizes $M^{l}M^{l\dagger}/M^{l\dagger}M^{l}$) involved in $V_{L/R}^{l(n)}$ (see Eq.(\ref{Vmatrix})). 

In Table \ref{tab:3l}, we show the values of the branching ratios for all possible FCNC lepton decay channels 
induced by exchanges of the first KK excitation of the $Z^0$ boson and photon (effects of higher KK states 
are discussed below). We have derived these values with $m_{KK}=1 \mbox{TeV}$ 
for the case Y in Appendix \ref{appendB}. We see in this table that all branching ratios are lower 
than their experimental upper limit, as wanted. Similarly, the branching ratios for the two other cases 
X and Z, given in appendix, also satisfy all relevant experimental bounds. In addition, we notice from 
Table \ref{tab:3l} that the amplitudes for the processes $\tau^- \to e^- \mu^+ \mu^-$ and $\tau^- \to e^+ \mu^- \mu^-$, 
for example, are different. The former involves the FC coupling $Z/\gamma^{(1)} \bar e \tau$ 
(fixed by $V^{l(1)}_{L/R13}$) whereas the latter involves $Z/\gamma^{(1)} \bar \mu \tau$ (fixed by $V^{l(1)}_{L/R23}$). 

\begin{table}[t] 
\begin{center} 
\begin{tabular}{c|c|c|c} 
Process &  $Z^{(1)}$    &    $\gamma^{(1)}$  &  Experimental 
\\ 
&     &     &  bound 
\\ 
\hline 
$B(\mu^- \to e^- e^+ e^-)$ &   $1.4 \ 10^{-14}$    &  $9.4 \ 10^{-14}$    & $1.0 \ 10^{-12}$ 
\\ 
\hline 
$B(\tau^- \to e^- e^+ e^-)$  &  $1.1 \ 10^{-12}$   &  $8.5 \ 10^{-12}$    & $2.9 \ 10^{-6}$ 
\\ 
\hline 
$B(\tau^- \to \mu^- \mu^+ \mu^-)$  &  $5.5 \ 10^{-8}$  &  $3.0 \ 10^{-7}$   & $1.9 \ 10^{-6}$ 
\\ 
\hline 
$B(\tau^- \to e^- \mu^+ \mu^-)$ &  $9.0 \ 10^{-13}$  &  $6.6 \ 10^{-12}$    & $1.8 \ 10^{-6}$ 
\\ 
\hline 
$B(\tau^- \to e^+ \mu^- \mu^-)$ & $1.5 \ 10^{-17}$   &  $9.7 \ 10^{-17}$    & $1.5 \ 10^{-6}$ 
\\ 
\hline 
$B(\tau^- \to \mu^- e^+ e^-)$ & $7.0 \ 10^{-8}$    &  $3.8 \ 10^{-7}$  & $1.7 \ 10^{-6}$ 
\\ 
\hline 
$B(\tau^- \to \mu^+ e^- e^-)$ &  $2.4 \ 10^{-22}$   & $2.1 \ 10^{-21}$     & $1.5 \ 10^{-6}$ 
\end{tabular} 
\caption{Branching ratios for all FCNC lepton decay channels induced by exchanges of the first 
KK excitations $Z^{(1)}$ and $\gamma^{(1)}$, for $m_{KK}=1 \mbox{TeV}$ and point Y of 
Appendix \ref{appendB}. The associated upper experimental bounds at $90\%\ C.L.$, taken from \cite{PDG04}, 
are also shown.} 
\label{tab:3l} 
\end{center} 
\end{table} 

In the present discussion on FCNC constraints, we do not present FCNC rates 
associated to exchanges of higher KK excitations ($n=2,\dots$) of gauge bosons $Z/\gamma^{(n)}$, 
as those are much weaker than the $Z/\gamma^{(1)}$ contributions to FCNC rates. The reason is 
that, compared to the $Z/\gamma^{(1)}$, the $Z/\gamma^{(n=2,\dots)}$ masses are larger and their 
absolute couplings to SM fermions, proportional to $|C^{(n)}(c_i)|$, are smaller whatever the 
fermion location parameter $c_i$ is (even getting smaller typically as the KK level $(n)$ increases) 
as shown clearly in \cite{HRizzoBIS}. 
For example for the same case Y as in Table \ref{tab:3l}, we find a branching ratio, 
of the decay channel $\tau^- \to \mu^- e^+ e^-$ (receiving the largest $Z/\gamma^{(1)}$ 
contributions) induced by the $Z^{(2)}$ ($\gamma^{(2)}$) exchange, equal to $1.2 \ 10^{-14}$ 
($6.2 \ 10^{-14}$) for an identical $m_{KK}=1 \mbox{TeV}$ which leads to 
$m_{KK}^{(2)}(\gamma)=(5.57/2.45) m_{KK}=2.27 \mbox{TeV}$.   

\vspace{0.5cm} 

$\bullet$ {\bf $Z^0 \to l_\alpha l_\beta$ decay:} 
The mixing between $Z^0$ boson and modes $Z^{(n)}$ gives rise to the FC lepton decay channels 
$Z^0 \to l_\alpha \bar l_\beta$ [$\alpha \neq \beta$]. From the general formalism described in 
\cite{Langacker}, for FC effects due to an additional heavy $Z'$ boson, one can easily deduce the   
width expressions for these leptonic decays in terms of matrix $V_{L/R\alpha\beta}^{(n)}$. We find the 
corresponding branching ratio values in Table \ref{tab:Zll}. This table shows that, for $m_{KK}=1 \mbox{TeV}$, 
the cases X,Y,Z do not conflict with the experimental bounds on rates of $Z^0$ FC decays into leptons. 
Finally, we note that for $m_{KK}=1 \mbox{TeV}$, the $Z^0-Z^{(1)}$ mixing angle is given by 
$\sin^2 \theta \approx 7 \ 10^{-5}$. 

\begin{table}[t] 
\begin{center} 
\begin{tabular}{c|c|c|c|c} 
Decay &  X   &   Y  &  Z & Experimental 
\\ 
channel &     &  &   &  bound       
\\ 
\hline 
$B(Z^0 \to e^\pm \mu^\mp)$ & $2.1 \ 10^{-17}$   &  $3.7 \ 10^{-15}$    & $3.2 \ 10^{-15}$ & $1.7 \ 10^{-6}$ 
\\ 
\hline 
$B(Z^0 \to e^\pm \tau^\mp)$ & $3.3 \ 10^{-15}$    &  $1.7 \ 10^{-12}$  & $1.8 \ 10^{-12}$ & $9.8 \ 10^{-6}$ 
\\ 
\hline 
$B(Z^0 \to \mu^\pm \tau^\mp)$ &  $2.1 \ 10^{-14}$   & $1.0 \ 10^{-7}$     & $7.6 \ 10^{-8}$ & $1.2 \ 10^{-5}$ 
\end{tabular} 
\caption{Branching ratios for FC leptonic decays induced by $Z^0-Z^{(1)}$ mixing, 
for $m_{KK}=1 \mbox{TeV}$ and the 3 points X,Y,Z of Appendix \ref{appendB}, with relevant 
upper experimental bounds at $95\%\ C.L.$ (from \cite{PDG04}).} 
\label{tab:Zll} 
\end{center} 
\end{table} 

\vspace{0.5cm} 

$\bullet$ {\bf $P^0-\bar P^0$ mixing:} 
Next, we study FCNC reactions in the hadron sector, starting by processes with $\Delta F=2$: 
The KK gauge field exchanges at tree level generate mass splittings in neutral pseudo-scalar meson systems. 
The mass splitting $\Delta m_P$ between flavour eigenstates for a meson $P$ was given in \cite{Langacker} 
as a function of meson decay constant $f_P$ and off-diagonal elements of matrix $V_{L/R}^{(n)}$. In 
Table \ref{tab:KKbar}, we present the values for mass splitting of the kaon induced by $Z^{(1)}$ and 
$\gamma^{(1)}$ exchanges: the results, obtained for $m_{KK}=1 \mbox{TeV}$, show that the 3 cases A,B,C 
of Appendix \ref{appendB} give contributions smaller than the experimental uncertainty which reads as 
$\Delta m_K=[3.483 \pm 0.006] \ 10^{-15} \ \mbox{GeV}$ \cite{PDG04}. 

\begin{table}[t] 
\begin{center} 
\begin{tabular}{c|c|c|c} 
Contribution &  A   &   B  &  C 
\\ 
\hline 
$\Delta m_K \ (Z^{(1)})$  & $1.1 \ 10^{-21}$   &  $1.7 \ 10^{-21}$    & $8.8 \ 10^{-21}$ 
\\ 
\hline 
$\Delta m_K \ (\gamma^{(1)})$  & $3.0 \ 10^{-23}$    &  $4.4 \ 10^{-23}$  & $2.3 \ 10^{-22}$ 
\end{tabular} 
\caption{Mass splitting (in $\mbox{GeV}$) for $K^0$ meson generated by $Z^{(1)}$ and $\gamma^{(1)}$, 
with $m_{KK}=1 \mbox{TeV}$ and the 3 points A,B,C of Appendix \ref{appendB}.} 
\label{tab:KKbar} 
\end{center} 
\end{table} 

Similarly, in Table \ref{tab:BBbar}, we give the values for mass splitting of the B meson. These 
values show that the 3 cases A,B,C lead to contributions which do not saturate the measured value: 
$\Delta m_B=[3.304 \pm 0.046] \ 10^{-13} \ \mbox{GeV}$ \cite{PDG04}. 

\begin{table}[t] 
\begin{center} 
\begin{tabular}{c|c|c|c} 
Contribution &  A   &   B  &  C 
\\ 
\hline 
$\Delta m_B \ (Z^{(1)})$  & $3.9 \ 10^{-18}$   &  $7.6 \ 10^{-18}$    & $1.6 \ 10^{-17}$ 
\\ 
\hline 
$\Delta m_B \ (\gamma^{(1)})$  & $1.0 \ 10^{-19}$    &  $2.0 \ 10^{-19}$  & $4.2 \ 10^{-19}$ 
\end{tabular} 
\caption{Mass splitting (in $\mbox{GeV}$) for $B^0$ meson generated by $Z^{(1)}$ and $\gamma^{(1)}$, 
with $m_{KK}=1 \mbox{TeV}$ and the 3 points A,B,C of Appendix \ref{appendB}.} 
\label{tab:BBbar} 
\end{center} 
\end{table} 

The mass splittings for D meson are given in Table \ref{tab:DDbar}. The values   
obtained for the cases A,B,C are in perfect agreement with the experimental limit,   
$\Delta m_D < 4.6 \ 10^{-14} \ \mbox{GeV}$ at $95\%\ C.L.$ \cite{PDG04}. 

\begin{table}[t] 
\begin{center} 
\begin{tabular}{c|c|c|c} 
Contribution &  A   &   B  &  C 
\\ 
\hline 
$\Delta m_D \ (Z^{(1)})$  & $5.5 \ 10^{-20}$   &  $1.9 \ 10^{-19}$    & $3.4 \ 10^{-19}$ 
\\ 
\hline 
$\Delta m_D \ (\gamma^{(1)})$  & $1.5 \ 10^{-21}$    &  $4.9 \ 10^{-21}$  & $8.8 \ 10^{-21}$ 
\end{tabular} 
\caption{Mass splitting (in $\mbox{GeV}$) for $D^0$ meson generated by $Z^{(1)}$ and $\gamma^{(1)}$, 
with $m_{KK}=1 \mbox{TeV}$ and the 3 points A,B,C of Appendix \ref{appendB}.} 
\label{tab:DDbar} 
\end{center} 
\end{table} 

Mass splittings in meson systems are also produced by exchanges of KK gluon excitations mediating 
FC. These $\Delta m_P$ contributions are larger than the excited EW gauge boson ones, due to the high 
strength of the strong interaction. However, considering for example the $B^0$ meson, the KK gluon contribution 
to mass splitting can remain well below the experimental error on $\Delta m_B$, as we are going to see. 
The first KK gluon contribution to the mass splitting for the $B^0$ meson, which has a mass (of $5279.4 \mbox{MeV}$) 
much larger than the QCD-scale \cite{Buras96}, can be estimated \cite{Burdman03} and yields respectively 
$\Delta m_B=\{1.5;3.0;6.2\} \times 10^{-16} \mbox{GeV}$ for cases A,B,C with $m_{KK}=1 \mbox{TeV}$. These 
values are about three orders of magnitude above the $\gamma^{(1)}$ contributions (see Table \ref{tab:BBbar}) 
but are still about one order below the experimental uncertainty on $\Delta m_B$ which is 
$\pm 4.6 \ 10^{-15} \mbox{GeV}$ (see text above). 

\vspace{0.5cm} 

$\bullet$ {\bf $\mu-e$ conversion:} 
Some FCNC reactions involve both quarks and leptons. The exchanges of neutral KK gauge fields mediating 
FC lead to coherent $\mu-e$ conversion in muonic atoms. The SINDRUM II collaboration at PSI has carried 
out a program of experiments to search for $\mu-e$ conversion in various nuclei \cite{SINweb} and the best 
exclusion limit obtained comes from the Titanium reaction \cite{Wintz}: 
\begin{equation} 
B(\mu^- + Ti \to e^- + Ti) 
=\frac{\Gamma (\mu^- + Ti \to e^- + Ti)}{\Gamma_{\mbox{\normalsize CAPT}}} 
<6.1 \ 10^{-13} \ \ \mbox{at} \ 90\% \ C.L., 
\label{TiConv} 
\end{equation} 
$\Gamma_{\mbox{\small CAPT}}$ being the total nuclear muon capture rate in $Ti$ 
which is measured with a good precision (see \cite{Bernabeu} for a nuclear-model-independent 
bound on the vertex $Z'-e-\mu$). The expression for the branching ratio 
$B(\mu^- + Ti \to e^- + Ti)$ can be deduced from global analysis in 
\cite{Langacker} where the FC amplitudes of an additional $Z'$ boson are calculated (taking into account 
the $Z^0-Z'$ mixing). One obtains this branching ratio as a function of a nuclear form factor, 
the Titanium atomic (Z=22) and neutron (N=26) numbers and the matrix elements $V^{l(1)}_{L/R12}$, 
$V^{u(1)}_{L/R11}$ and $V^{d(1)}_{L/R11}$. We find that the value of this branching ratio respects 
the bound in Eq.(\ref{TiConv}) for any set of quark parameters A,B,C combined with any set X,Y,Z for 
leptons and with $m_{KK}=1 \mbox{TeV}$, as can be checked from Table \ref{tab:mueConv} where values 
are given for examples of combinations. 

\begin{table}[t] 
\begin{center} 
\begin{tabular}{c|c|c|c} 
Contribution &  A/X   &   B/Y  &  C/Z 
\\ 
\hline 
$B(Z^{(1)})$  & $9.5 \ 10^{-16}$   & $7.5 \ 10^{-14}$  &  $4.7 \ 10^{-14}$   
\\ 
\hline 
$B(\gamma^{(1)})$  & $1.6 \ 10^{-15}$   & $4.8 \ 10^{-14}$ &  $1.5 \ 10^{-14}$   
\end{tabular} 
\caption{Branching fraction for $\mu^- + Ti \to e^- + Ti$ induced by $Z^{(1)}$ or 
$\gamma^{(1)}$ exchange, with $m_{KK}=1 \mbox{TeV}$ and for 3 representative 
combinations of the points listed in Appendix \ref{appendB}.} 
\label{tab:mueConv} 
\end{center} 
\end{table} 

\vspace{0.5cm} 

$\bullet$ {\bf $K^0 \to l_\alpha l_\beta$ decay:} 
Tight experimental constraints apply on certain (semi-) leptonic FCNC decay amplitudes for mesons. 
First, the decay channels of type $K^0_L \to l_\alpha \bar l_\beta$ receive contributions from 
processes involving KK gauge boson excitations. The $K^0_L$ branching fraction associated to such 
contributions is directly derived from the general results of the systematic survey of lowest-dimension 
effective interactions (as manifestations of heavy physics) performed in \cite{London}: we obtain 
the branching ratio $B(K^0_L \to l_\alpha \bar l_\beta)$ as function of the $B(K^+ \to \nu_\mu \bar \mu)$ 
value and the matrix elements $V^{d(1)}_{L/R12}$, $V^{l(1)}_{L/R\alpha\beta}$. In our framework, the computed 
values of $B(K^0_L \to l_\alpha \bar l_\beta)$ are in agreement with the associated experimental values 
(either much smaller than the measurement error or lower than the existing bound) for all sets of quark 
parameters in the cases A,B,C combined with any set in X,Y,Z for leptons and for $m_{KK}=1 \mbox{TeV}$, 
as can be observed in Table \ref{tab:Kll} in which values are presented for some characteristic examples 
of combinations. 

\begin{table}[t] 
\begin{center} 
\begin{tabular}{c|c|c|c|c} 
Branching &  A/X   &   B/Y  &  C/Z & Experimental 
\\ 
ratio  &  &  &   &  value 
\\ 
\hline 
$B(K_L \to e^+ e^-)$  & 
$4.9 \ 10^{-14}$ & 
$1.1 \ 10^{-14}$ &   
$1.7 \ 10^{-14}$ & 
$[9^{+6}_{-4}] \ 10^{-12}$     
\\ 
& 
$3.6 \ 10^{-13}$  & 
$8.0 \ 10^{-14}$  &   
$1.3 \ 10^{-13}$  & 
\\ 
\hline 
$B(K_L \to \mu^+ \mu^-)$  & 
$4.9 \ 10^{-14}$ & 
$8.5 \ 10^{-15}$ &   
$1.1 \ 10^{-14}$ & 
$[7.27  \pm 0.14] \ 10^{-9}$     
\\ 
& 
$3.6 \ 10^{-13}$  & 
$6.2 \ 10^{-14}$  &   
$8.4 \ 10^{-14}$  & 
\\ 
\hline 
$B(K_L \to e^\pm \mu^\mp)$  & 
$1.7 \ 10^{-26}$ & 
$4.5 \ 10^{-24}$ &   
$2.1 \ 10^{-23}$ & 
$< \ 4.7 \ 10^{-12}$ 
\\ 
& 
$1.7 \ 10^{-25}$  & 
$4.0 \ 10^{-23}$  &   
$1.9 \ 10^{-22}$  & 
\end{tabular} 
\caption{Branching ratio of decays $K_L \to l_\alpha \bar l_\beta$ for $m_{KK}=1 \mbox{TeV}$ and the 3 
combinations A/X,B/Y,C/Z of points given in Appendix \ref{appendB}. 
The first number corresponds to the $Z^{(1)}$ 
contribution and the second one is for $\gamma^{(1)}$. In last column, we also provide each measured 
branching fraction value with its uncertainty, as well as the experimental upper limit at $90\%\ C.L.$ 
in the case of the FC final state $e^\pm \mu^\mp$ \cite{PDG04}.} 
\label{tab:Kll} 
\end{center} 
\end{table} 

\vspace{0.5cm} 

$\bullet$ {\bf $K^+ \to \pi^+ \nu \nu$ decay:} 
The exchange of $Z^{(1)}$ contributes to the $K^+ \to \pi^+ \nu \bar \nu$ FCNC decay 
(with an implicit summation over the indexes $\alpha,\beta$ of final state neutrinos 
$\nu_\alpha \bar \nu_\beta$ including $\alpha \neq \beta$ channels). 
The SM contribution to this decay leads to a branching fraction of 
$B_{SM}=(0.4 \ \mbox{to} \ 1.2) \ 10^{-10}$ in agreement with the experimental result: 
$B_{exp}=[1.6^{+1.8}_{-0.8}] \ 10^{-10}$ \cite{PDG04}. Hence, the maximal allowed value 
for the ratio $B_{RS}/B_{SM}$ ($B_{RS}$ representing the branching fraction of 
$K^+ \to \pi^+ \nu \bar \nu$ induced in the RS model by the $Z^{(1)}$ exchange) is 
typically between $1.8$ and $3.6$. This ratio $B_{RS}/B_{SM}$ can be expressed 
\cite{Buras96,Burdman} in terms of the matrix elements $V^{d(1)}_{L/R12}$, 
$V^{\nu(1)}_{L\alpha\beta}$. 
We find that the $B_{RS}/B_{SM}$ value is clearly below the limit discussed above, 
for all the combinations of A,B,C with X,Y,Z and $m_{KK}=1 \mbox{TeV}$. In Table 
\ref{tab:KPinunu}, we give the $B_{RS}/B_{SM}$ value for the same examples of parameter 
combinations as before.   

\begin{table}[t] 
\begin{center} 
\begin{tabular}{c|c|c|c} 
Contribution &  A/X   &   B/Y  &  C/Z 
\\ 
\hline 
$R(K^+ \to \pi^+ \nu \bar \nu)$  & $6.7 \ 10^{-6}$ & $1.1 \ 10^{-6}$ &  $1.3 \ 10^{-6}$     
\end{tabular} 
\caption{Ratio $R=B_{RS}/B_{SM}$ for the decay channel $K^+ \to \pi^+ \nu \bar \nu$ 
(induced only by the $Z^{(1)}$ exchange in case of the RS model) for $m_{KK}=1 \mbox{TeV}$ 
and the same 3 combinations of points given by Appendix \ref{appendB}.} 
\label{tab:KPinunu} 
\end{center} 
\end{table} 

\vspace{0.5cm} 

$\bullet $ \textbf{$l_{\alpha }\rightarrow l_{\beta }\gamma $ decay:} One 
loop neutral current penguin diagrams, exchanging a $Z/\gamma ^{(1)}$ and a 
charged lepton, induce FC radiative decays into the photon: $l_{\alpha 
}\rightarrow l_{\beta }\gamma $ [$\alpha \neq \beta $]. The amplitudes of 
such diagrams can be expressed, using the formalism in \cite{Langacker}, in 
terms of the matrices $V_{L/R}^{l(1)}$. In Table \ref{tab:llgamma}, the 
branching ratios of various decay channels $l_{\alpha }\rightarrow l_{\beta 
}\gamma $ are given for $m_{KK}=1\mbox{TeV}$ and the 3 lepton parameter sets 
X,Y,Z: we see that none of those values conflicts with experimental bounds. 

\begin{table}[t] 
\begin{center} 
\begin{tabular}{c|c|c|c|c} 
Branching & X & Y & Z & Experimental \\ 
ratio &  &  &  & limit \\ \hline 
$B(\mu^- \to e^- \gamma)$ & $1.1 \ 10^{-13}$ & $1.4 \ 10^{-13}$ & $1.1 \ 
10^{-13}$ & $1.2 \ 10^{-11}$ \\ 
& $8.3 \ 10^{-13}$ & $1.0 \ 10^{-12}$ & $8.3 \ 10^{-13}$ &  \\ \hline 
$B(\tau^- \to e^- \gamma)$ & $1.4 \ 10^{-14}$ & $2.9 \ 10^{-14}$ & $1.4 \ 
10^{-14}$ & $2.7 \ 10^{-6}$ \\ 
& $3.3 \ 10^{-14}$ & $1.1 \ 10^{-13}$ & $3.3 \ 10^{-14}$ &  \\ \hline 
$B(\tau^- \to \mu^- \gamma)$ & $4.9 \ 10^{-11}$ & $6.8 \ 10^{-11}$ & $4.9 \ 
10^{-11}$ & $1.1 \ 10^{-6}$ \\ 
& $2.1 \ 10^{-11}$ & $3.0 \ 10^{-11}$ & $2.1 \ 10^{-11}$ & 
\end{tabular} 
\end{center} 
\caption{Branching ratio for all decays $l_\protect\protect\alpha \to 
l_\protect\protect\beta \protect\gamma$ [$\protect\alpha \neq \protect\beta$] 
for $m_{KK}=1 \mbox{TeV}$ and the 3 points X,Y,Z of Appendix \protect\ref{appendB}. 
First value is associated to the $Z^{(1)}$ contribution while the 
second one is for $\protect\gamma^{(1)}$. In last column, we give the 
experimental upper limit at $90\%\ C.L.$ for each decay channel \protect\cite{PDG04}.} 
\label{tab:llgamma} 
\end{table} 

\vspace{0.3cm} 

Next, we discuss another type of contribution to the FC decay channel 
$l_{\alpha }\rightarrow l_{\beta }\gamma $ [$\alpha \neq \beta $]. In the SM 
extension where neutrinos have (Dirac) masses, this radiative decay is 
mediated by the exchange of a $W^{\pm }$ gauge boson and a neutrino at one 
loop-level. In this case, the source of FC resides in the lepton mixing 
matrix $V_{MNS}=U_{L}^{l\dagger }U_{L}^{\nu }$. Within this scenario, the 
rate for the FC decay $l_{\alpha }\rightarrow l_{\beta }\gamma $ is 
suppressed by the GIM cancellation mechanism \cite{GIM} which is ensured, 
simultaneously, by the unitarity of $V_{MNS}$ and the quasi-degeneracy of 
the 3 neutrino masses (relatively to the $W^{\pm }$ mass). \newline 
In the RS model with bulk matter, loop contributions of KK neutrino 
excitations \cite{HSpre} to $l_{\alpha }\rightarrow l_{\beta }\gamma $ 
invalidate \cite{Ilakovac} the GIM cancellation. Indeed, these excitations have KK 
masses which are not negligible (and thus not quasi-degenerate in family space) 
compared to $m_{W^{\pm }}$. The GIM mechanism is also invalidated by the 
loop contributions of the KK $W^{\pm (n)}$ modes which couple (KK level by 
level), in the 4-dimensional theory, via an effective lepton mixing matrix 
of type $V_{MNS}^{eff}=U_{L}^{l\dagger} \mathcal{C}_{L}^{(n)} U_{L}^{\nu}$ 
being non-unitary due to the non-universality of 
$\mathcal{C}_{L}^{(n)}=diag(C_{m}^{(n)}(c_{1}^{L}),C_{m}^{(n)}(c_{2}^{L}),C_{m}^{(n)}(c_{3}^{L}))$, 
where $m\geq 0$ is the exchanged neutrino KK level index and 
$C_{m}^{(n)}=C_{0mn}^{f\bar{f}A}$ (in the notation of \cite{HRizzoBIS}). 
\newline 
However, for the cases considered in Appendix \ref{appendB}, the three 
$c_{i}^{L}$ values are equal (for case X) or almost equal (for Y and Z). 
Remember that only the $c_{i}^{L}$ values play a r\^{o}le here, as the 
leptons coupling to $W^{\pm }$ must be left handed. Thus, the GIM 
cancellation mechanism is restored for these cases, KK level by level \cite{Song}, 
in the process $l_{\alpha }\rightarrow l_{\beta }\gamma $. Indeed, the two 
arguments given above do not hold anymore. 
First, for three (almost) equal $c_{i}^{L}$ values, the 3 family 
masses of the excited neutrino states at a common KK level, exchanged in the 
loop, are (quasi-)degenerate with respect to $m_{W^{\pm }}$ \cite{HRizzoBIS}. 
Secondly, for (almost) identical $c_{i}^{L}$'s, the effective 
matrix $V_{MNS}^{eff}=U_{L}^{l\dagger} \mathcal{C}_{L}^{(n)} U_{L}^{\nu}$ of KK 
$W^{\pm (n)}$ modes (almost) verifies $V_{MNS}^{eff}V_{MNS}^{eff\dagger     
}\propto {1\>\!\!\!\mathrm{I}}_{3\times 3}$ since: \textit{(i)} 
$\mathcal{C}_{L}^{(n)}=diag(C_{m}^{(n)}(c_{1}^{L}),C_{m}^{(n)}(c_{2}^{L}),C_{m}^{(n)}(c_{3}^{L}))$ 
is (quasi-)universal \textit{(ii)} $U_{L}^{\nu }$ is nearly unitary (m)-KK 
level by level, as the neutrino masses are much smaller than their KK 
excitation masses, i.e. no significant mixings are induced among different 
neutrino KK levels. We conclude that for the parameter space described in 
Appendix \ref{appendB}, the dominant contributions to the widths of FC decay 
$l_{\alpha }\rightarrow l_{\beta }\gamma $ must originate from the exchanges 
of KK neutral gauge fields discussed before (see also the next discussion on 
$b\rightarrow s\gamma $). 

\vspace{0.5cm} 

$\bullet $ \textbf{$b\rightarrow s\gamma $ decay:} Similarly, one loop 
penguin diagrams, exchanging a neutral KK gauge field and a down quark, 
contribute to the FC radiative partonic decay: $b\rightarrow s\gamma $. The 
experimental measurement of $R_{b\rightarrow s\gamma }$ yields \cite{PDG04}, 
\begin{equation} 
R_{b\rightarrow s\gamma }\equiv \frac{\Gamma (B\rightarrow X_{s}\gamma )}{\Gamma 
(B\rightarrow X_{c}e\bar{\nu}_{e})}=3.39_{-0.54}^{+0.62}\ 10^{-3}, 
\label{ExpBSG} 
\end{equation} 
where the $\Gamma $'s denote the widths. For the SM expectation, one has 
$R_{b\rightarrow s\gamma }^{SM}=3.23\pm 0.09\ 10^{-3}$ \cite{PDG04}. 
Therefore, the contribution of the KK gauge fields to $R_{b\rightarrow 
s\gamma }$ cannot exceed $8.7\ 10^{-4}$. The $b$ quark mass is much higher 
than the QCD-scale. Thus, long-range strong interaction effects are not 
expected to be important in the decay $B\rightarrow X_{s}\gamma $ 
\cite{Buras96}. Hence, the ratio $R_{b\rightarrow s\gamma }$ is usually 
approximated by, 
\begin{equation} 
R_{b\rightarrow s\gamma }\simeq \frac{\Gamma (b\rightarrow s\gamma )}{\Gamma 
(b\rightarrow ce\bar{\nu}_{e})},  \label{AppBSG} 
\end{equation} 
The expression for this ratio, as a function of the $V_{L/R}^{d(1)}$ matrix 
elements and phase-space factors, can be easily deduced 
from previous study \cite{Langacker}: the values of the KK gauge field 
contributions to the ratio in Eq.(\ref{AppBSG}), obtained for quark parameter 
sets A,B,C and $m_{KK}=1\mbox{TeV}$, are much smaller than the experimental 
bound discussed above, as exhibits Table \ref{tab:bsgamma}. 

\begin{table}[t] 
\begin{center} 
\begin{tabular}{c|c|c|c} 
Contribution & A & B & C \\ \hline 
$R_{b \to s \gamma} \ (Z^{(1)})$ & $9.4 \ 10^{-16}$ & $7.0 \ 10^{-16}$ & 
$2.6 \ 10^{-16}$ \\ \hline 
$R_{b \to s \gamma} \ (\gamma^{(1)})$ & $1.8 \ 10^{-16}$ & $8.0 \ 10^{-17}$ 
& $1.9 \ 10^{-17}$ 
\end{tabular} 
\end{center} 
\caption{Contributions to the ratio $R_{b \to s \protect\gamma}$ (see text) 
from $Z^{(1)}$ and $\protect\gamma^{(1)}$ exchanges, for $m_{KK}=1 \mbox{TeV} 
$ and the 3 points A,B,C listed in Appendix \protect\ref{appendB}.} 
\label{tab:bsgamma} 
\end{table} 

\vspace{0.3cm} 

We finish this part by discussing the contribution to $b\rightarrow s\gamma $ 
coming from the exchange of a $W^{\pm (n)}$ [$n=0,1\dots $] gauge field and 
an up quark (or its KK excitations) at one loop-level. Analogous arguments 
as those employed in the discussion on $l_{\alpha }\rightarrow 
l_{\beta }\gamma$ apply here. Hence, the major contributions to the $b\rightarrow s\gamma $ 
rate should not come from the $W^{\pm (n)}$ exchange for the cases 
considered here, where all $c_{i}^{Q}$ values are exactly equal between all 
families (\textit{c.f.} Appendix \ref{appendB}). \newline 
Nevertheless, in the quark sector, there are deviations to the restoration 
of the GIM cancellation discussed above, due to the fact that the top quark 
mass cannot be totally neglected relatively to the KK up-type quark 
excitation scales. Indeed, this fact leads to a mass shift of the KK top 
quark mode from the rest of the KK up-type quark modes, and thus eliminates 
the degeneracy among 3 family masses of the up quark excitations at fixed KK 
level (with regard to $m_{W^{\pm (n)}}$). Moreover, this means that the 
Yukawa interaction with the Higgs boson induces a substantial mixing of 
the top quark KK tower members among themselves \cite{Victoria,Kaplan}. 
\newline 
As an example, the data on $b\rightarrow s\gamma $ are accommodated with 
$m_{KK}\simeq 1\mbox{TeV}$ for $c_{3}^{Q}=0.4$ (which is close to the values in 
the cases B and C of Appendix \ref{appendB}), as shown in \cite{Song} using 
numerical methods for the diagonalization of a large dimensional mass matrix and 
taking into account the top quark mass effects described previously. 

\section{Other constraints} 

\label{other} 

\subsection{EW precision data} 

\label{EWmeas} 

EW precision measurements place restrictions on the RS model 
\cite{HRizzoBIS,Burdman,Song,EWboundA}-\cite{EWboundD} (with bulk matter) since 
deviations from EW precision observables arise in the framework of this 
model. Indeed, the mixing between the top quark and its KK excitations 
(discussed above) results in a shift of the ratio $m_{W^{\pm }}/m_{Z^{0}}$ 
from the value obtained within the SM. Moreover, mixing 
between the EW gauge bosons and their KK modes induces modifications 
of the boson masses/couplings. The authors in \cite{EWboundA} have found that a 
good fit of EW precision observables, including the $\rho $ parameter, can 
be obtained with $m_{KK}\approx 11\mbox{TeV}$. In \cite{HRizzoBIS}, a global 
analysis based on a large set of EW observables has yielded a lower bound on 
$m_{KK}$ varying typically between $0$ and $20\mbox{TeV}$ for a universal 
value $|c_{i}|<1$. If the weak gauge boson masses and couplings are treated 
simultaneously, one obtains the conservative bound $m_{KK}\gtrsim 10\mbox{TeV},$ 
for a universal $c_{i}$ value lying inside the interval $[-1,1]$ 
(and for $10^{-2}<k/M_{5}<1$) \cite{EWboundC}. 

\vspace{0.3cm} 

Different specific models have been proposed in literature in order to soften this 
lower bound on $m_{KK}$ coming from EW precision data. The motivation was to 
address the little hierarchy problem, i.e. the smallness of the EW 
symmetry breaking scale compared to $m_{KK}$. In this sense, the EW bound on 
$m_{KK}$ is model-dependent. For example, models with brane-localized 
kinetic terms for fermions \cite{BraneF} or gauge bosons \cite{BraneB} allow 
to relax the lower bound on $m_{KK}$ down to a few TeV (see \cite{EWBa,EWBb} 
for gauge boson kinetic terms and \cite{EWF} for fermion terms). The 
introduction of brane-localized kinetic terms changes the KK wave functions 
so that our results, presented here, cannot be directly translated to such models 
by a simple rescaling with the appropriate powers of $m_{KK}$. A different type 
of model, with a left-right EW gauge structure in the bulk and already 
mentioned \cite{LR}, also allows for softening the EW bound on $m_{KK}$, 
thanks to the bulk custodial isospin gauge symmetry arising in this framework. 
Our realizations of the RS scenario with bulk matter can be considered within the 
context of this type of models, allowing to combine all EW precision data with 
a value for $m_{KK}\gtrsim 3\mbox{TeV}$. We discuss this in the 
next paragraph. 

\vspace{0.3cm} 

In the model of \cite{LR}\footnote{The 
Higgs phenomenology in left-right symmetric RS extensions was analysed 
in \cite{Ben}.}, with the EW gauge symmetry enhanced to $SU(2)_{L}\times 
SU(2)_{R}\times U(1)_{B-L}$, the usual EW gauge group $SU(2)_{L}\times 
U(1)_{Y}$ is recovered through the breaking of both $SU(2)_{R}$ and 
$U(1)_{B-L}$ on the Planck-brane (scenario II) and possibly a small 
breaking of $SU(2)_{R}$ in the bulk (scenario I). Furthermore, the right 
handed fermions are promoted to $SU(2)_{R}$ doublet fields, with the new 
(non physical) component having no zero mode. Hence, for instance in the 
quark sector, the right handed $c_{i}^{d,u}$ parameters describe now the 
locations of $SU(2)_{R}$ doublets, however, the total number of 
$c_{i}^{Q,d,u}$ parameters remains identical. Let us discuss the possibility, 
within this context, that our types of configurations for $c_{i}$ parameters 
are in agreement with a reasonable fit of the EW precision data for 
$m_{KK}\simeq 3\mbox{TeV}$. \newline 
$\bullet $ \textbf{$\delta g_{Z}$ shift:} The coupling $g_{Z}^{b}$ of 
the $Z^{0}$ boson to the $b$ quark is measured with high accuracy. 
Experimentally, this is done through the width ratio $R_{b}=\Gamma 
(b\bar{b})/\Gamma (hadronic)$ \cite{PDG04}. For $m_{KK}=3\mbox{TeV}$ and 
$c_{3}^{Q}=0.200;0.370;0.413$, corresponding to the cases A;B;C 
respectively, the shift in the coupling (obtained from formula 5.9 in 
\cite{LR}) is $\delta g_{Z}^{b_{L}}/g_{Z}^{b_{L}} \approx 1.7\%;0.8\%;0.6\%$ which 
respects the upper limit on $\delta g_{Z}^b/g_{Z}^b$ 
of the order of the percent as imposed in \cite{LR}. 
The left handed $b_{L}$ quark (with $c_{3}^{Q}<0.5$ as dictated by the top quark 
mass) is considered, since its effective couplings to KK gauge 
fields are much larger than the couplings of the $b_{R}$ quark (with 
$c_{3}^{d}>0.5,$ systematically). Our $c_{i}^{L}$ values are 
also lower than $0.5$. These values lead respectively to the 
shift amounts $\delta g_{Z}^{l_{L}}/g_{Z}^{l_{L}} \approx 3.2\%;1.1\%;0.4\%$ 
for $c^{L}=-1.50;0.20;0.39$ in cases X;Y;Z, if $m_{KK}=4\mbox{TeV}$   
(using results of \cite{LR} with the $Q_Z$ and $Q_{Z'}$ charges for charged leptons). 
The ratios $R_{e}$, $R_{\mu}$ and $R_{\tau}$ give rise to precisions, on 
the $Z^{0}$ couplings $g_{Z}^{l}$ (not $g_{Z}^{l_{L}}$) 
to charged leptons, which are of the same 
order as those on $g_{Z}^{b}$. Besides, we remark that the calculation of shift 
in the couplings performed in \cite{LR} does not strictly take into account the SM 
fermion mixing angles: these mixing angles enter the couplings between SM fermions and 
KK gauge boson excitations (as in Eq.(\ref{Sphysical})-(\ref{Vmatrix})) 
via the unitary matrices diagonalizing 
the mass matrices (which are fixed by the precise values of $c_i$ and Yukawa couplings). 
These mixings could reduce significantly the couplings with KK states, and thus the 
deviations of EW observables from their SM value. 
\newline 
$\bullet $ \textbf{S parameter:} The value of the \textquotedblleft 
oblique\textquotedblright\ parameter S is found \cite{LR} to be typically 
$0.26-0.15$ for $m_{KK}=3-4\mbox{TeV}$ (when $c_{3}^{Q},c_{3}^{u}<0.5$ and 
$c_{3}^{d}>0.5$). For lower $m_{KK}$ values, S is too large in order to 
reasonably fit EW precision data (independently of T). \newline 
$\bullet $ \textbf{T parameter:} In scenario I (mentioned above), for 
$kR_{c}\sim 10$, $m_{KK}\simeq 3\mbox{TeV}$ and the above range of S, the T 
parameter reaches values required to fit the EW measurements \cite{LR}. 
One notes that the $c_i$ values in our case C, for instance, are close 
to configurations of $c_i$ considered in \cite{LR}. 
In scenario II, the correct T values required for above S range can be 
generated radiatively from top loops. Indeed, using expression 6.4 of 
\cite{LR}, we find $T_{\mbox{KK \ top}}\simeq 0.15$ for $c_{3}^{u}\approx 0$, 
$c_{3}^{Q}=0.37$ (as in our case B) and $m_{KK}=3\mbox{TeV}$ (the involved 
$m_{KK}^{(1)}(t_{L})$ mass being fixed by $c_{3}^{Q}$ and $m_{KK}$). 

\subsection{Universality limits} 

\label{Univ} 

Especially for low KK masses, mixing between the zero modes of SM fermions 
and their KK excitations induce a loss of flavour universality for the 
effective quark/lepton couplings to neutral gauge bosons. Indeed, the 
existence of these mixings causes a loss of unitarity (in zero mode fermion 
flavour space) with regard to the matrices responsible for the 
transformation from weak to physical basis. The largest deviation, induced 
by such effects, from the SM value of the fermion couplings to the $Z^{0}$ 
boson arises for the top quark. 

Under the hypothesis that the LHC measures the top coupling to the $Z^{0}$ 
with a precision of $5\%$ (an accuracy of a few percent is expected to be 
reached in the LHC performances) \textit{and} that the result coincides with 
the SM prediction, an experimental lower limit could be placed on the mass 
$m_{KK}^{(1)}(t)$ of first KK top quark excitation. This hypothetical limit, 
obtained in \cite{Santiago}, is $m_{KK}^{(1)}(t)\gtrsim 1-4\mbox{TeV}$ for a 
universal $c_{i}$ value in the range $[0,0.5]$, corresponding to a less 
severe bound on $m_{KK}$ which is systematically smaller than 
$m_{KK}^{(1)}(t)$. The other indirect constraints of this type, not involving 
the top quark, are less restrictive. 

\subsection{Muon magnetic moment} 

\label{Anom} 

The anomalous magnetic moment of the muon is a well known model building 
constraint on theories beyond SM. In the RS framework, this magnetic moment 
receives contributions from the loop exchanges of KK excitations. The 
experimental world average measurement of $(g-2)_{\mu }$ translates into the 
upper limit $c_{i}\lesssim 0.70$ for $1\mbox{TeV}<m_{KK}<10\mbox{TeV}$, 
assuming a universal value for all the $c_{i}$ parameters \cite{Muon}. 
Because of this simplification assumption, i.e. $c_{i}$ universality, this 
upper limit does not strictly apply to our realistic RS scenario, where the 
values of $c_{i}$ parameters are flavour and type dependent. 

The authors of \cite{Muon} have also examined the perturbativity condition 
on effective Yukawa coupling constants from which they have deduced the constraint 
$c_{i}\lesssim 0.77$, still under the hypothesis of a universal $c_{i}$ value. 

\section{Non-renormalizable operators} 

\label{NR} 

In models of low gravity scale yielding a low cutoff, the impact of 
non-renormalizable operators is dramatically amplified. This constitutes a 
serious challenge for model building. Within the RS framework, the 
fundamental value of 5-dimensional gravity scale in the bulk (where SM 
fields propagate, in the present RS set-up) $M_{5}$ is close to the high 
$M_{Pl}$ value (\textit{c.f.} Eq.(\ref{RSkrelat})). However, one should ask 
whether effective 4-dimensional non-renormalizable interactions, determined 
by field overlaps along the fifth dimension, are sufficiently suppressed. 

Let us, explicitly, express the effective 4-dimensional energy scale 
($Q_{\alpha \beta \gamma \delta}$) 
of four fermion operators, relevant for FC reactions, in the physical basis. 
We start from the generic four fermion operator in the fundamental theory, 
assuming $M_{5}$ as the characteristic energy scale and taking all dimensionless 
coupling constants $\lambda_{ijkl}$ equal to unity: 
\begin{equation} 
\int d^{4}x\int dy\ \sqrt{G}\ \frac{\lambda_{ijkl}}{M_{5}^{3}}\ \bar{\Psi}_{i}\Psi _{j} 
\bar{\Psi}_{k}\Psi _{l}=\int d^{4}x\ \frac{1}{Q_{\alpha \beta \gamma \delta}^{2}}\ \bar{\psi}_{\alpha 
}^{(0)\ \prime }\psi _{\beta }^{(0)\ \prime }\bar{\psi}_{\gamma }^{(0)\ 
\prime }\psi _{\delta }^{(0)\ \prime }+\dots   
\label{eq:NRfund} 
\end{equation} 
We remind that $_{i,j,k,l}$ are family indexes of the weak basis and 
$_{\alpha ,\beta ,\gamma ,\delta }$ flavour indexes of the mass basis. The 
dots stand for KK excitation coupling terms. The expression for the effective 
energy scale $Q_{\alpha \beta \gamma \delta}$ in mass basis, obtained after the 
integration over $y $ and using Eq.(\ref{RSkrelat}), reads as 
(with an implicit sum over $i,j,k,l$) 
\begin{equation} 
\frac{1}{Q_{\alpha \beta \gamma \delta}^{2}}= 
\frac{U_{\alpha i}^{\dagger }U_{j\beta }U_{\gamma k}^{\dagger }U_{l\delta }} 
{\Lambda_{ijkl}^2} 
\label{eq:NRscale} 
\end{equation} 
where the matrices $U_{i\alpha }$ are responsible for the basis transformation of SM 
fermions (see Section \ref{origin}) and 
the 4-dimensional energy scale $\Lambda_{ijkl}$ is given by 
\begin{equation} 
\frac{1}{\Lambda_{ijkl}^{2}}=\frac{\lambda_{ijkl}}{N_{0}^{i}N_{0}^{j}N_{0}^{k}N_{0}^{l}}\ 
\frac{1-e^{-2\pi kR_{c}}}{2\pi ^{2}(kR_{c})^{2}M_{Pl}^{2}}\ 
\frac{e^{\pi k R_{c}(4-c_{i}-c_{j}-c_{k}-c_{l})}-1}{4-c_{i}-c_{j}-c_{k}-c_{l}}. 
\label{eq:NRscaleFIN} 
\end{equation} 
The normalization factors $N_{0}^{i}$ were defined by Eq.(\ref{Norm}). 

In the following subsections, we calculate the effective scale 
$Q_{\alpha \beta \gamma \delta}$, numerically, 
for the various types of four fermion operators contributing to FCNC processes: we 
will show that, for the 6 sets of parameters given in Appendix \ref{appendB} 
which fix both the $U_{i\alpha }$ matrices and $N_{0}^{i}$ factors (and thus 
$Q_{\alpha \beta \gamma \delta}$), 
the obtained $Q_{\alpha \beta \gamma \delta}$ values induce different FCNC effects 
respecting all associated experimental constraints. 

\subsection{Lepton FC decays} 

\label{NRlept} 

Some four fermion operators induce the leptonic three-body decays of FCNC type 
$l_{\alpha }\rightarrow l_{\beta }l_{\gamma }l_{\delta }$, at a rate given 
approximately by $\Gamma \simeq m_{l_{\alpha }}^{5}/Q_{\alpha \beta \gamma \delta}^{4}$ 
(omitting the phase space factors) as deduced from Eq.(\ref{eq:NRfund}). In Table 
\ref{tab:NR3l}, we explicitly present all the kinds of higher dimensional 
operators contributing to such decay channels. We restrict ourselves to 
operators which originate from EW gauge invariant terms and have allowed chirality 
configurations. These operators are written in terms of zero mode fields in 
the mass basis. For each one of these operators, we show, in the same table, 
the numerical value of the associated scale $Q_{\alpha \beta \gamma \delta}$ obtained from our 
theoretical expression in Eq.(\ref{eq:NRscale})-(\ref{eq:NRscaleFIN}) for the point Y of parameter 
space. Our conclusion, here, is that all effective $Q_{\alpha \beta \gamma \delta}$ scale values 
obtained are well above their experimental lower limit. Indeed, the 
constraint on the branching ratio $B(\mu ^{-}\rightarrow 
e^{-}e^{+}e^{-})<1.0\ 10^{-12}$ ($B(\tau ^{-}\rightarrow 
l_{\beta }l_{\gamma}l_{\delta })\lesssim 2\ 10^{-6}$) \cite{PDG04}, considered previously in 
Table \ref{tab:3l}, translates into an experimental limit $Q_{2111}>2.6\ 
10^{6}\mbox{GeV}$ ($Q_{3 \beta \gamma \delta}\gtrsim 5\ 10^{4}\mbox{GeV}$). The same results 
hold for the X and Z cases, i.e. the $Q_{\alpha \beta \gamma \delta}$ values systematically 
satisfy these experimental limits. 

\begin{table}[t] 
\begin{center} 
\begin{tabular}{c|c|c|c} 
Decay channel & Operator & $\bar l_L^c l_L \bar l_L l_L^c$ & $\bar l_R^c l_R 
\bar l_R l_R^c$ \\ \hline 
$\mu^- \to e^- e^+ e^-$ & $\bar \mu^c e \bar e e^c$ & $2.0 \ 10^{12}$ & $6.1 
\ 10^{12}$ \\ \hline 
$\tau^- \to e^- e^+ e^-$ & $\bar \tau^c e \bar e e^c$ & $9.4 \ 10^{10}$ & $1.5 \ 10^{12}$ \\ \hline 
$\tau^- \to \mu^- \mu^+ \mu^-$ & $\bar \tau^c \mu \bar \mu \mu^c$ & $4.1 \ 
10^{7}$ & $2.6 \ 10^{11}$ \\ \hline 
$\tau^- \to e^- \mu^+ \mu^-$ & $\bar \tau^c \mu \bar \mu e^c$ & $7.4 \ 
10^{8} $ & $6.8 \ 10^{11}$ \\ \hline 
$\tau^- \to e^+ \mu^- \mu^-$ & $\bar \tau^c e \bar \mu \mu^c$ & $7.4 \ 
10^{8} $ & $6.8 \ 10^{11}$ \\ \hline 
$\tau^- \to \mu^- e^+ e^-$ & $\bar \tau^c e \bar \mu e^c$ & $1.7 \ 10^{10}$ 
& $1.1 \ 10^{12}$ \\ \hline 
$\tau^- \to \mu^+ e^- e^-$ & $\bar \tau^c \mu \bar e e^c$ & $1.7 \ 10^{10}$ 
& $1.1 \ 10^{12}$ 
\end{tabular} 
\end{center} 
\caption{Four fermion operator types inducing the different FCNC lepton 
decays, together with their associated effective 4-dimensional $Q_{\alpha \beta \gamma \delta}$ scale 
value (in units of $\mbox{GeV}$) for the point Y of Appendix \protect\ref{appendB}. 
This value is given for the two possible chirality configurations 
of each operator, as indicated in the last two columns. We use conventions 
for Dirac spinors meaning that chirality projection acts first, then charge 
conjugation second and Dirac bar third: $\bar l_{L/R}^c=\overline{(l_{L/R})^c}$.} 
\label{tab:NR3l} 
\end{table} 

\subsection{Meson mass splittings} 

\label{NRqua} 

Other types of four fermion operators contribute to the mass splitting in 
neutral pseudo-scalar meson systems. In Table \ref{tab:NRsplit}, we give the 
gauge invariant forms, allowed by chirality, of dimension-six operators 
contributing to the $\Delta m_{K}$ mass splitting (in terms of the zero mode 
quarks in the physical basis). For each operator, we also give the value of 
the corresponding scale $Q_{1122}$ for the 3 sets A,B,C. With regard to this 
table, we observe that the values found for $Q_{1122}$ satisfy the experimental 
bound $Q_{1122}>5\ 10^{6}\mbox{GeV}$ \cite{RSloc,Hflav} imposed by constraints 
on $K^{0}-\bar{K}^{0}$ mixing (studied in Section \ref{small}). 

\begin{table}[t] 
\begin{center} 
\begin{tabular}{c|c|c|c} 
Operator & A & B & C \\ \hline & & & \\ 
$\bar d_L^c d_L \bar s_L s_L^c$ & $5.9 \ 10^{7}$ & $1.2 \ 10^{8}$ & $1.7 \ 
10^{8}$ \\ \hline & & & \\ 
$\bar d_R^c d_R \bar s_R s_R^c$ & $4.1 \ 10^{12}$ & $2.1 \ 10^{12}$ & $1.3 \ 
10^{12}$ 
\end{tabular} 
\end{center} 
\caption{$Q_{1122}$ energy scale (in $\mbox{GeV}$) of the operators contributing 
to $\Delta m_K$ for the 3 points A,B,C of Appendix \protect\ref{appendB}. 
Recall that in our spinorial notation, one has for the down quark: $\bar 
d_{L/R}^c=\overline{(d_{L/R})^c}$.} 
\label{tab:NRsplit} 
\end{table} 

Similarly, for the cases A,B,C, the values obtained for the $Q_{1133}$ scale of 
the $(\bar{d}^{c}d)(\bar{b}b^{c})$ operator contributing to $\Delta m_{B}$ 
are respectively equal to $2.4\ 10^{6}$, $4.8\ 10^{6}$, $7.0\ 10^{6}\mbox{GeV}$ 
for left handed states, and $1.3\ 10^{11}$, $6.6\ 10^{10}$, $4.0\ 10^{10}\mbox{GeV}$ 
for the right handed ones. These values are all within (although close to, 
for the left handed states) their experimental bound: $Q_{1133}>2\ 10^{6}\mbox{GeV}$. 
\newline 
The 3 cases A,B,C also give rise to $Q_{2211}$ values, for the operator 
$(\bar{c}^{c}c)(\bar{u}u^{c})$, which are clearly in agreement with the experimental 
constraints coming from $D^{0}-\bar{D}^{0}$ mixing. 

\subsection{Muon electron conversion} 

\label{NRleptoquark} 

Certain non-renormalizable operators involving both quarks and leptons can 
lead to $\mu -e$ conversion in muonic atoms. Indeed, the operators presented 
in Table \ref{tab:NRleptoq} generate this conversion. On this table, we also show 
the corresponding 
effective $Q_{1112}$ values computed for three characteristic combinations 
involving the quark parameter sets A,B,C and the lepton sets X,Y,Z. One can 
check there that the values obtained for $Q_{1112}$ are well within the 
experimental constraint $Q_{1112}>10^{5}\mbox{GeV}$ originating from the 
exclusion limit on $B(\mu ^{-}+Ti\rightarrow e^{-}+Ti)$ discussed in 
Eq.(\ref{TiConv}). In fact, any combination of A,B,C together with X,Y,Z leads to 
acceptable values for the $Q_{1112}$ mass scale. 

\begin{table}[t] 
\begin{center} 
\begin{tabular}{c|c|c|c} 
Operator & A/X & B/Y & C/Z \\ \hline & & & \\   
$\bar e_R d_L \bar d_L \mu_R$ & $1.7 \ 10^{11}$ & $7.2 \ 10^{10}$ & $6.8 \ 
10^{10}$ \\ \hline & & & \\ 
$\bar e_L d_R \bar d_R \mu_L$ & $1.3 \ 10^{11}$ & $8.2 \ 10^{11}$ & $9.3 \ 
10^{11}$ \\ \hline & & & \\ 
$\bar e_R u_L \bar u_L \mu_R$ & $3.6 \ 10^{11}$ & $1.6 \ 10^{11}$ & $1.5 \ 
10^{11}$ \\ \hline & & & \\ 
$\bar e_L u_R \bar u_R \mu_L$ & $2.6 \ 10^{9}$ & $1.2 \ 10^{11}$ & $1.7 \ 
10^{11}$ 
\end{tabular} 
\end{center} 
\caption{Effective scale $Q_{1112}$ (in $\mbox{GeV}$) of the four operators 
contributing to coherent $\protect\mu - e$ conversion, for 3 combinations of 
sets A,B,C and X,Y,Z taken from Appendix \protect\ref{appendB}.} 
\label{tab:NRleptoq} 
\end{table} 

\section{Conclusion} 

\label{conclu} 

>From the study on the RS model (with bulk matter) presented here, we obtain 
the following main conclusion. Regardless of the details of the model, 
there exist certain types of configuration for the fermion localizations which, 
simultaneously, reproduce the fermion mass hierarchies and mixings, and, 
generate FCNC effects within the present experimental limits for low KK 
gauge boson masses. The impact of this conclusion is important, for two 
reasons: \newline 
First, this new possibility of the existence of light KK gauge boson states 
constitutes one of the first motivations for experimental searches of gauge 
boson excitations at the next coming high energy colliders 
\footnote{Most of the previous phenomenological works on RS model signatures at 
colliders, found in the literature, were dedicated to processes exchanging 
the KK excitations of graviton in the original RS set-up with all SM fields 
trapped at the TeV-brane.}. \newline 
Second, the possibility of having low KK gauge boson masses allows for a 
good solution of the gauge hierarchy problem within the RS scenario. As a 
matter of fact, low $m_{KK}$ masses permit large values for $kR_{c}$ and 
thus small $M_{\star }$ gravity scale 
values, close to the EW scale 
\footnote{For example, with $m_{KK}=1\mbox{TeV}$ (in 
agreement with FCNC constraints, as we have shown), the theoretical 
condition (\ref{Rbound}) on $k$ dictates the maximum $kR_{c}$ value to be 
$10.83$ leading to $M_{\star }=4\mbox{TeV}$, which is almost of the same 
order as the EW scale.}. 

\vspace{0.3cm} 

In a detailed analysis, we have constructed complete realizations of 
the RS scenario that address the gauge hierarchy 
problem, reproduce all the present data on quark/lepton masses and mixing 
angles (for the case of Dirac neutrinos), induce FCNC process amplitudes 
satisfying the experimental bounds for KK masses down to $m_{KK}=1\mbox{TeV}$ 
and generate acceptable effective suppression scales for non-renormalizable 
operators in the physical basis (for the parameter product in Eq.(\ref{kRvalue})). 
It seems that our types of configurations for fermion locations are potentially 
compatible with some RS extensions suggested in the literature, respecting EW 
precision constraints with $m_{KK}\gtrsim 3\mbox{TeV}$. Nevertheless, a detailed 
study is required. 

\vspace{0.3cm} 

In other words, we have shown that the attractive version of the RS model, 
providing a geometrical interpretation for the huge SM fermion mass hierarchies, 
does not necessarily 
conflict with the existence of small KK gauge field masses around $3\mbox{TeV}$. 
Thus, in particular, it should induce diverse characteristic signatures potentially 
detectable at LHC. Indeed, even if a precise experimental investigation would 
be needed to prove the feasibility of such a detection, a preliminary study performed 
in \cite{HRizzoBIS}, under the simplification assumption of a unique 
universal $c_{i}$ value 
(which clearly prevents the creation of quark/lepton mass hierarchies), 
already obtained the following indicative results: the Tevatron Run II (with an integrated 
luminosity of $\mathcal{L}=2fb^{-1}$) is capable of testing masses up to 
$m_{KK}\simeq 1\mbox{TeV}$ via a direct search for the first KK excited gauge 
boson exchanges, while the expected LHC sensitivity (for 
$\mathcal{L}=100fb^{-1}$) on $m_{KK}$ can reach values up to about $6\mbox{TeV}$. 

\vspace{1cm} 

\noindent \textbf{\Large Acknowledgments} 

\noindent The authors thank K.~Agashe, G.~Bhattacharyya, G.~C.~Branco, S.~J.~Huber and 
P.~F.~P\'erez for useful conversations. G.~M. acknowledges support from a 
\textit{Marie Curie} Intra-European Fellowships (under contract MEIF-CT-2004-514138) 
within the 6th European Community Framework Program. 

\newpage 

\appendix 
\noindent \textbf{\Large Appendix} \vspace{0.5cm} 

\renewcommand{\thesubsection}{A.\arabic{subsection}} 
\renewcommand{\theequation}{A.\arabic{equation}} 
\setcounter{subsection}{0} 
\setcounter{equation}{0} 

\section{Experimental data} 
\label{appendA} 

At the $Z^0$ boson mass scale $Q=m_{Z^0}$, the renormalized charged 
lepton masses are \cite{Fusaoka} 
\begin{equation} 
\begin{array}{l} 
m_e=0.48684727 \pm 1.4 \ 10^{-7} \ \mbox{MeV} 
\\ 
m_\mu=102.75138 \pm 3.3 \ 10^{-4} \ \mbox{MeV} 
\\ 
m_\tau=1.74669^{+0.00030}_{-0.00027} \ \mbox{GeV} 
\end{array} 
\label{dataL} 
\end{equation} 
the quark masses are \cite{Fusaoka} 
\begin{equation} 
\begin{array}{ll} 
m_d=4.69^{+0.60}_{-0.66} \ \mbox{MeV} \ ;  &  m_u=2.33^{+0.42}_{-0.45} \ \mbox{MeV}   
\\ 
m_s=93.4^{+11.8}_{-13.0} \ \mbox{MeV} \ ;  &  m_c=677^{+56}_{-61} \ \mbox{MeV}   
\\ 
m_b=3.00 \pm 0.11 \ \mbox{GeV} \ ;  &  m_t=181 \pm 13 \ \mbox{GeV}   
\end{array} 
\label{dataQ} 
\end{equation} 
and the three CKM matrix parameters are \cite{Fusaoka} 
\begin{equation} 
\begin{array}{rcl} 
|V_{us}| & = & 0.2205 \pm 0.0018  \\ 
|V_{cb}| & = & 0.0373 \pm 0.0018  \\ 
|V_{ub}/V_{cb}| & = & 0.08 \pm 0.02 . 
\end{array} 
\label{dataV} 
\end{equation}

Next, we give the current data on neutrino masses and lepton mixings. 
A general three-flavour fit to the world's global neutrino data sample has been 
performed in \cite{Valle}: the data sample used in this analysis includes the 
results from solar, atmospheric, reactor (KamLAND and CHOOZ) and accelerator (K2K) 
experiments. The values for oscillation parameters obtained in this analysis 
at the $4\sigma$ level are contained in the intervals 
\begin{equation*} 
6.8\leq \Delta m_{21}^{2}\leq 9.3\ \ \ [10^{-5}\mbox{eV}^{2}], 
\end{equation*} 
\begin{equation} 
1.1\leq \Delta m_{31}^{2}\leq 3.7\ \ \ [10^{-3}\mbox{eV}^{2}] 
\label{4SigmaDataA} 
\end{equation} 
where $\Delta m_{21}^{2}\equiv m_{\nu _{2}}^{2}-m_{\nu _{1}}^{2}$ and 
$\Delta m_{31}^{2}\equiv m_{\nu _{3}}^{2}-m_{\nu _{1}}^{2}$ are the 
differences of squared neutrino mass eigenvalues, and 
\begin{equation*} 
0.21\leq \sin ^{2}\theta _{12}\leq 0.41, 
\end{equation*} 
\begin{equation*} 
0.30\leq \sin ^{2}\theta _{23}\leq 0.72, 
\end{equation*} 
\begin{equation} 
\sin ^{2}\theta _{13}\leq 0.073   
\label{4SigmaDataB} 
\end{equation} 
where $\theta _{12}$, $\theta _{23}$ and $\theta _{13}$ are the three mixing 
angles of the convenient form of parameterization for the lepton mixing 
matrix (MNS matrix) now adopted as standard by the Particle Data 
Group \cite{PDG04}. 
Furthermore, the data on tritium beta decay \cite{107} provided by the Mainz 
\cite{108} and Troitsk \cite{109} experiments give rise to the following upper 
bounds at $95\%\ C.L.$, 
\begin{equation*} 
m_{\beta }\leq 2.2\ \mbox{eV}\ \ \ \mbox{[Mainz]}, 
\end{equation*} 
\begin{equation} 
m_{\beta }\leq 2.5\ \mbox{eV}\ \ \ \mbox{[Troitsk]}  \label{mbetaLIM} 
\end{equation} 
with the effective mass $m_{\beta}$ defined by $m_{\beta}^{2}= 
\sum_{i=1}^{3}|U_{ei}|^{2}m_{\nu_{i}}^{2}$, $U_{ei}$ denoting 
the lepton mixing matrix elements. 

\renewcommand{\thesubsection}{B.\arabic{subsection}} 
\renewcommand{\theequation}{B.\arabic{equation}} 
\setcounter{subsection}{0} 
\setcounter{equation}{0} 

\section{Points of parameter space} 
\label{appendB} 

We give here 3 complete sets [A,B,C] 
of parameters, namely all Yukawa coupling constants and 5-dimensional masses 
(see Section \ref{values} for notations and conventions), reproducing the present quark masses 
and mixing angles (which are summarized in Appendix \ref{appendA}): 
\vskip 0.5cm 
\begin{equation*} 
[A] 
\end{equation*} 
\begin{equation*} 
\begin{array}{lll} 
c_{1}^{Q}=0.2\ ;\ \  & c_{2}^{Q}=0.2\ ;\ \  & c_{3}^{Q}=0.2 
\\ 
c_{1}^{d}=0.728\ ;\ \  & c_{2}^{d}=0.740\ ;\ \  & c_{3}^{d}=0.628 
\\ 
c_{1}^{u}=0.62\ ;\ \  & c_{2}^{u}=0.62\ ;\  & c_{3}^{u}=0.35 
\end{array} 
\end{equation*} 
\begin{equation*} 
\kappa_{ij}^{d}= 
\left ( 
\begin{array}{ccc} 
1.0 & 1.0 & 1.01 
\\   
1.1 & -0.9 & 0.952 
\\   
1.0 & 1.0 & 1.067 
\end{array} 
\right ) 
\ \ \ 
\kappa_{ij}^{u}= 
\left ( 
\begin{array}{ccc} 
1.0 & 0.9 & 1.03 
\\   
1.1 & 1.0 & 0.9 
\\   
1.0 & 0.9 & 1.1 
\end{array} 
\right ) 
\end{equation*} 
\vskip 0.5cm 
\begin{equation*} 
[B] 
\end{equation*} 
\begin{equation*} 
\begin{array}{lll} 
c_{1}^{Q}=0.37\ ;\ \  & c_{2}^{Q}=0.37\ ;\ \  & c_{3}^{Q}=0.37 
\\ 
c_{1}^{d}=0.716\ ;\ \  & c_{2}^{d}=0.728\ ;\ \  & c_{3}^{d}=0.615 
\\ 
c_{1}^{u}=0.607\ ;\ \  & c_{2}^{u}=0.607\ ;\  & c_{3}^{u}=0.050 
\end{array} 
\end{equation*} 
\begin{equation*} 
\kappa_{ij}^{d}= 
\left ( 
\begin{array}{ccc} 
1.0 & 1.0 & 1.017 
\\   
1.1 & -0.9 & 0.96 
\\   
1.0 & 1.0 & 1.075 
\end{array} 
\right ) 
\ \ \ 
\kappa_{ij}^{u}= 
\left ( 
\begin{array}{ccc} 
1.0 & 0.9 & 1.029 
\\   
1.1 & 1.0 & 0.9 
\\   
1.0 & 0.9 & 1.1 
\end{array} 
\right ) 
\end{equation*} 
\vskip 0.5cm 
\begin{equation*} 
[C] 
\end{equation*} 
\begin{equation*} 
\begin{array}{lll} 
c_{1}^{Q}=0.413\ ;\ \  & c_{2}^{Q}=0.413\ ;\ \  & c_{3}^{Q}=0.413 
\\ 
c_{1}^{d}=0.703\ ;\ \  & c_{2}^{d}=0.721\ ;\ \  & c_{3}^{d}=0.608 
\\ 
c_{1}^{u}=0.6\ ;\ \  & c_{2}^{u}=0.6\ ;\  & c_{3}^{u}=-0.4 
\end{array} 
\end{equation*} 
\begin{equation*} 
\kappa_{ij}^{d}= 
\left ( 
\begin{array}{ccc} 
1.0 & 1.0 & 1.017 
\\   
1.1 & -0.9 & 0.96 
\\   
1.0 & 1.0 & 1.075 
\end{array} 
\right ) 
\ \ \ 
\kappa_{ij}^{u}= 
\left ( 
\begin{array}{ccc} 
1.0 & 0.9 & 1.029 
\\   
1.1 & 1.0 & 0.9 
\\   
1.0 & 0.9 & 1.1 
\end{array} 
\right ) 
\end{equation*} 
\vskip 0.5cm 
the Yukawa coupling indexes $i$ and $j$ corresponding respectively to the line and column, 
exactly as in Eq.(\ref{Yuk})-(\ref{MassMatrix}). 

Now, we present 3 sets [X,Y,Z] of parameters creating the current data on lepton masses and mixings ({\it c.f.} 
Appendix \ref{appendA}): 
\vskip 0.5cm 
\begin{equation*} 
[X] 
\end{equation*} 
\begin{equation*} 
\begin{array}{lll} 
c_{1}^{L}=-1.5\ ;\ \  & c_{2}^{L}=-1.5\ ;\ \  & c_{3}^{L}=-1.5 
\\ 
c_{1}^{l}=0.760\ ;\ \  & c_{2}^{l}=0.833\ ;\ \  & c_{3}^{l}=0.667 
\\ 
c_{1}^{\nu}=1.512\ ;\ \  & c_{2}^{\nu}=1.513\ ;\  & c_{3}^{\nu}=1.468 
\end{array} 
\end{equation*} 
\begin{equation*} 
\kappa_{ij}^{l}= 
\left ( 
\begin{array}{ccc} 
0.9 & 1.0 & 1.1 
\\   
1.0 & 1.1 & 1.1 
\\   
-1.1 & 0.9 & 0.9 
\end{array} 
\right ) 
\ \ \ 
\kappa_{ij}^{\nu}= 
\left ( 
\begin{array}{ccc} 
-1.1 & -0.9 & -1.1 
\\   
-1.1 & 1.0 & -1.1 
\\   
-0.9 & 0.9 & 0.9 
\end{array} 
\right ) 
\end{equation*} 
\vskip 0.5cm 
\begin{equation*} 
[Y] 
\end{equation*} 
\begin{equation*} 
\begin{array}{lll} 
c_{1}^{L}=0.200\ ;\ \  & c_{2}^{L}=0.200\ ;\ \  & c_{3}^{L}=0.261 
\\ 
c_{1}^{l}=0.737\ ;\ \  & c_{2}^{l}=0.696\ ;\ \  & c_{3}^{l}=0.647 
\\ 
c_{1}^{\nu}=1.496\ ;\ \  & c_{2}^{\nu}=1.503\ ;\  & c_{3}^{\nu}=1.463 
\end{array} 
\end{equation*} 
\begin{equation*} 
\kappa_{ij}^{l}= 
\left ( 
\begin{array}{ccc} 
1.0 & 1.0 & 1.0 
\\   
1.0 & 1.003 & 1.0 
\\   
-0.9 & 1.0 & 1.0 
\end{array} 
\right ) 
\ \ \ 
\kappa_{ij}^{\nu}= 
\left ( 
\begin{array}{ccc} 
-1.0 & -1.1 & -1.0 
\\   
-1.1 & 1.0 & -1.1 
\\   
-1.0 & 1.0 & 0.9 
\end{array} 
\right ) 
\end{equation*} 
\vskip 0.5cm 
\begin{equation*} 
[Z] 
\end{equation*} 
\begin{equation*} 
\begin{array}{lll} 
c_{1}^{L}=0.35\ ;\ \  & c_{2}^{L}=0.35\ ;\ \  & c_{3}^{L}=0.39 
\\ 
c_{1}^{l}=0.728\ ;\ \  & c_{2}^{l}=0.694\ ;\ \  & c_{3}^{l}=0.636 
\\ 
c_{1}^{\nu}=1.49\ ;\ \  & c_{2}^{\nu}=1.49\ ;\  & c_{3}^{\nu}=1.45 
\end{array} 
\end{equation*} 
\begin{equation*} 
\kappa_{ij}^{l}= 
\left ( 
\begin{array}{ccc} 
1.0 & 1.0 & 1.0 
\\   
1.0 & 1.0035 & 1.0 
\\   
-0.9 & 1.0 & 1.0 
\end{array} 
\right ) 
\ \ \ 
\kappa_{ij}^{\nu}= 
\left ( 
\begin{array}{ccc} 
-1.0 & -1.1 & -1.0 
\\   
-1.1 & 1.0 & -1.1 
\\   
-1.0 & 1.0 & 0.9 
\end{array} 
\right ) 
\end{equation*}

\end{document}